\documentclass[12pt,a4paper]{iopart}
\usepackage[T1]{fontenc}
\usepackage[latin9]{inputenc}
\usepackage{framed, enumerate}
\usepackage[margin=2cm]{geometry}
\usepackage{graphicx}
\usepackage{amssymb}
\usepackage{esint,url,bm}

\makeatletter

\usepackage{amsthm}\usepackage{amsfonts}\usepackage{epsf}\usepackage{epstopdf}\usepackage{latexsym}\usepackage{framed}\@ifundefined{definecolor}
 {\usepackage{color}}{}

\newcommand{\tfrac}[2]{\frac{#1}{#2}}

\newcommand{\sgn}{\mbox{sgn}}
\newcommand{\bfi}{\bfseries\itshape}

\newcommand{\rem}[1]{}



\usepackage[english]{babel}

\begin{document}

\title[{Two-component cross-coupled CH2}]{Waltzing peakons and compacton pairs in a cross-coupled Camassa-Holm equation}
\author{
Colin J Cotter$^{1}$, Darryl D Holm$^{2}$, Rossen I Ivanov$^{2,3}$
and James R Percival$^{2}$
}
\address{$^1$ Department of Aeronautics, Imperial College London, London SW7 2AZ, UK.}
\address{$^2$ Department of Mathematics, Imperial College London, London SW7 2AZ, UK.}
\address{$^3$ School of Mathematical Sciences, Dublin Institute of Technology, Kevin
Street, Dublin 8, Ireland,}

\eads{\mailto{colin.cotter@ic.ac.uk}, \mailto{d.holm@ic.ac.uk},
 \mailto{r.ivanov@imperial.ac.uk}, \mailto{rivanov@dit.ie}, \mailto{j.percival@ic.ac.uk}}

\date{{\it\color{red} Fondly remembering our late friend Jerry Marsden} }



\noindent \textbf{AMS Classification:} 

\noindent \textbf{Keywords:}


\begin{abstract}

We consider singular solutions of a system of two cross-coupled Camassa-Holm (CCCH) equations. This CCCH system admits peakon solutions, but it is not in the two-component CH integrable hierarchy. The
system is a pair of coupled Hamiltonian partial differential equations for two types of solutions on the real line, 
each of which separately possesses $\exp(-|x|)$ peakon solutions with a discontinuity in the first derivative
at the peak. However, there are no self-interactions, so each of the two types of peakon solutions 
moves only under the induced velocity of the other type. We analyse the `waltzing' solution behaviour of the cases with a single bound peakon pair (a peakon couple), as well as the over-taking collisions of peakon couples and the antisymmetric case of the head-on collision of a peakon couple and a peakon anti-couple. 
We then present numerical solutions of these collisions, which are \emph{inelastic} because the 
waltzing peakon couples each possess an internal degree of freedom corresponding to their `tempo' -- that is, the  period at which the two peakons of opposite type in the couple cycle around each other in phase space. Finally, we discuss compacton couple solutions of the cross-coupled Euler-Poincar\'e (CCEP) equations and illustrate the same types of collisions as for peakon couples, with triangular and parabolic compacton couples.
We finish with a number of outstanding questions and challenges remaining for understanding 
couple dynamics of the CCCH and CCEP equations. 




\end{abstract}


\section{Introduction}

\subsection{Purpose and scope of the paper} 
In this paper we investigate solution properties of a new two-component Camassa-Holm (CH) system
introduced in equations  (\ref{Xfleq1}) and  (\ref{Xfleq2}). As far as we know, this system is not completely integrable. However, the system does possess bound pairs of peakon solutions which exhibit interesting propagation dynamics involving both propagation and oscillation, while a single peakon must remain stationary (fixed in space). The oscillating and translating motion of the bound pairs of peakons as they propagate is reminiscent of the swirling motion of waltzing dancers, so we call these solutions {\it peakon couples}. The variational derivation of the equations, their geometry and the behaviour of the peakon couple propagation and collision interactions are studied in the paper. We also discuss compacton couple solutions for the more general case of cross-coupled Euler-Poincar\'e (CCEP) equations in (\ref{Xfleq1}) and  (\ref{Xfleq2}), and illustrate the same types of collisions as for peakon couples, with triangular and parabolic compacton couples.

\paragraph{CH equation and generalizations}

The well-known CH equation may be written as \cite{CH93,CHH94}
\begin{equation}
u_{t}-u_{xxt}+2\omega u_{x}+3uu_{x}-2u_{x}u_{xx}-uu_{xxx}=0,
\label{CH-eqn}
\end{equation}
where subscripts denote partial derivatives of the horizontal fluid velocity $u(x,t)$ and the linear dispersion parameter $\omega$ is a real constant. The solution $u(x,\,\cdot\,)$ and its derivative $u_x(x,\,\cdot\,)$ are assumed to vanish sufficiently rapidly at spatial infinity, 
$\lim_{|x|\to\infty}(u,u_x)(x,\,\cdot\,)=0$.
 
The integrable CH equation (\ref{CH-eqn}) describes the unidirectional propagation of shallow water waves
over a flat bottom at one order in asymptotic expansion beyond the KdV equation
\cite{CH93,CHH94,DGH03,DGH04,J02,J03a,CoLa2009}. 
The CH equation also governs axially symmetric waves in a hyperelastic rod \cite{Dai98}.
In the case of no linear dispersion, $\omega=0$, the CH equation (\ref{CH-eqn}) possesses singular solutions
in the form of peaked travelling waves called \emph{peakons} 
\begin{equation}
u(x,t) = 
\frac{c}{2}\exp({-|x-ct|})
\,.
\label{CH-peakon}
\end{equation}

Peakon properties and solution behaviour are reviewed in \cite{Ho2010}. Recent developments
about the CH equation may also be found in \cite{HoScSt2009,HI10}
and the references therein.

The scope of mathematical interpretations of CH may
be gleaned by rewriting it in various equivalent forms, many of which
have been rediscovered several times and each of which can be
a point of departure for further investigation. For example, CH may
be treated variously as: a fluid motion equation; a mathematical model
of shallow water wave breaking; a transport equation for wave momentum
by a fluid velocity related
to it by inversion of the Helmholtz operator; a nonlocal characteristic
equation; an Euler-Poincar\'e equation describing geodesic motion on
the diffeomorphism group with respect to the metric defined by the
$H^{1}$ norm on the tangent space of vector fields; a Lie-Poisson
Hamiltonian system describing coadjoint motion on the Bott-Virasoro
Lie group; a bi-Hamiltonian system; a compatibility equation for a
linear system of two equations in a Lax pair, etc.

In particular, the CH equation on the real line follows from Hamilton's principle $\delta{S}=0$ with an
action $S=\int l(u)dt$ given by Lagrangian
\begin{eqnarray}
l(u) = \frac12\int u^2+2\omega{u} + u_x^2 \,dx 
\label{CH-Lag}
\end{eqnarray}
for solutions $u(x,t)$ that vanish sufficiently rapidly at spatial infinity. 
The Euler-Poincar\'e equation \cite{HoMaRa1998}
\begin{eqnarray}
{\partial_t}m
=
-\,
{\rm ad}^*_um
=
-\,(um)_{x}-mu_x
\quad\hbox{with}\quad
m
:=
\frac{\delta l}{\delta u} = u - u_{xx} + \omega
,
\label{EP-eqn}
\end{eqnarray}
then recovers the CH equation (\ref{CH-eqn}).

The present paper emerged from the observation that \emph{complexifying} the CH Lagrangian in (\ref{CH-Lag}) yields
\begin{eqnarray}
l(u,v) &=& \frac12 \int (u+iv)^2 + 2\omega(u+iv) +  (u_x+iv_x)^2 \, dx
\nonumber\\
 &=& \tfrac12 \int (u^2 + u_x^2) + 2\omega{u}+ (v^2 + v_x^2)\, dx + i  \int uv + u_xv_x + \omega{v} \, dx
\\
 &=&
\Re(l) + i\, \Im(l) .
 \nonumber
\end{eqnarray}
Now the Euler-Poincar\'e equation (\ref{EP-eqn}) for the real part of the Lagrangian $\Re(l)$ simply yields two separate copies of the CH equation, one with $\omega $ and one without. This, of course, is of no further interest, because it gives nothing new. However, the Euler-Poincar\'e equation for the \emph{imaginary} part of the Lagrangian $\Im(l)$ yields a new coupled system of two equations for $u$ and $v$ that we shall investigate in the remainder of the present paper. This system of cross-coupled CH equations (CCCH) is found to be (with details of the derivation given in Section \ref{Sec 1})
\begin{eqnarray}
{\partial_t}m
&=&
-\,
{\rm ad}^*_v m
=
-\,(vm)_{x}-mv_x
\quad\hbox{with}\quad
m
:=
\frac{\delta l}{\delta v} = u - u_{xx}  + \omega
,
\label{CC-m-eqn}
\\
{\partial_t}n
&=&
-\,
{\rm ad}^*_u n
=
-\,(un)_{x}-nu_x
\quad\hbox{with}\quad
n
:=
\frac{\delta l}{\delta u} = v - v_{xx}
.
\label{CC-n-eqn}
\end{eqnarray}
%
%

In these equations the momentum $m$ (resp. $n$) is transported \emph{only} by the opposite induced velocity $v$ (resp. $u$).
Hereafter, we ignore linear dispersion by setting set $\omega =0$. This has the effect that: (i) the system is symmetric and (ii) it admits peakon solutions. Peakon solutions for the CCCH system (\ref{CC-m-eqn}) and (\ref{CC-n-eqn}) with $\omega =0$ are singular solutions in which the values of the momenta $m$ and $n$ are supported with time-dependent weights on delta functions at points moving along the real line, carried by the appropriate velocity. 

The present paper focuses its attention on the dispersionless cross-coupled CH system (CCCH) comprising (\ref{CC-m-eqn}) and  (\ref{CC-n-eqn}) with $\omega=0$. These equations govern \emph{two} types of peakons.
Each type of peakon is swept along by the induced flow of other peakons of the {\it opposite} type. 
We also consider solutions of the cross-coupled Euler-Poincar\'e equations CCEP, 
\begin{eqnarray}
{\partial_t}m
&=&
-\,
{\rm ad}^*_v m
=
-\,(vm)_{x}-mv_x
\quad\hbox{with}\quad
u=K_\mathcal{G}*m,
\label{CCEP-m-eqn}
\\
{\partial_t}n
&=&
-\,
{\rm ad}^*_u n
=
-\,(un)_{x}-nu_x
\quad\hbox{with}\quad
v=K_\mathcal{G}*n.
\label{CCEP-n-eqn}
\end{eqnarray}
For a convolution kernel $K_\mathcal{G}(\,\cdot\,)$ with compact spatial support, these CCEP equations admit \emph{compacton} solutions%
\footnote{The term `compacton' was introduced in \cite{RoHy1993} to describe solutions of nonlinear evolutionary partial differential equations that propagate coherently, interact essentially elastically and have compact support. }
whose scattering interactions we shall compute. 

To finish establishing the notation, the next paragraph will quickly review the well-known behaviour of the CH 
equation by writing in several of its various forms and discussing its properties that will be 
relevant here, in parallel with the properties of the CCCH system (\ref{CC-m-eqn}) and (\ref{CC-n-eqn}). We should mention that there are other, integrable two and multi-component CH systems with application in fluid mechanics, see e.g. \cite{OR96,CLZ05,I06,ELY07,CI08,HLT09,I09,HIFT,HI11}.

\subsection{Parallel properties for CH and CCCH}

\paragraph{Momentum conservation in CH and CCCH.}

The CH equation (\ref{CH-eqn}) with $\omega = 0$ 
\begin{equation}
u_{t}-u_{xxt} +3uu_{x}-2u_{x}u_{xx}-uu_{xxx}=0,
\label{CH-eqn-omega0}
\end{equation}
may be rewritten in the form of a {\bfi momentum conservation law},
as 
\begin{eqnarray}
u_{t}+\partial_{x}\left( \tfrac12 u^2 + P\right)=0
\,,
\end{eqnarray}
 with {\bfi pressure} $P$ given by the convolution 
 \begin{eqnarray}
P = K* \left(u^{2}+\tfrac{1}{2}u_{x}^{2} \right)
\quad\hbox{with kernel}\quad K(x,y)
= \tfrac{1}{2}\exp(-|x-y|).
\label{kernelK}
\end{eqnarray}
 The kernel $K$ for peakons is the Green's function for the 1D Helmholtz operator,
$(1-\partial_{x}^{2})$, and is also the shape of the peakon velocity
profile in (\ref{CH-peakon}).

When expressed in terms of the corresponding velocities $u=K*m$ and $v=K*n$, with kernel $K(x)$ given  by the Green's function in equation (\ref{kernelK}), the cross-coupled CH (CCCH) equations (\ref{CC-m-eqn}) and  (\ref{CC-n-eqn})  with $\omega=0$ expand into
\begin{eqnarray}
u_{t}-u_{xxt}+vu_{x}+2v_{x}u-2v_{x}u_{xx}-vu_{xxx} & = & 0,\label{Xfleq1}\\
v_{t}-v_{xxt}+uv_{x}+2u_{x}v-2u_{x}v_{xx}-uv_{xxx} & = & 0,\label{Xfleq2}
\end{eqnarray}

As for CH, the total momentum of CCCH is conserved, namely
\begin{eqnarray}
\partial_{t}\left(u+v\right)+\partial_{x}\big(uv+K*(2uv+u_{x}v_{x}) \big)=0.
\end{eqnarray}

\paragraph{Characteristic forms of CH and CCCH.}

The CH equation (\ref{CH-eqn-omega0}) may be interpreted as the condition for the 1-form
density $m\, dx^{2}$ to be preserved (or `frozen') in the
flow of the characteristic velocity $dx/dt=u(x(t),t)$, namely, 
\begin{eqnarray}
\frac{d}{dt}\Big(m\, dx^{2}\Big)=0\,,\quad\hbox{along}\quad\frac{dx}{dt}=u(x(t),t)=K*m.
\label{Lag-invCH}
\end{eqnarray}
Note that the relation between the velocity of the flow $u$ and the property that 
it carries $m$ is \emph{nonlocal}. 

Likewise, for the CCCH equations, but with the cross velocities, we have
\begin{eqnarray}
\frac{d}{dt}\Big(m\, dx^{2}\Big)&=&0\,,\quad\hbox{along}\quad\frac{dx}{dt}=v(x(t),t)=K*n,
\\
\frac{d}{dt}\Big(n\, dx^{2}\Big)&=&0\,,\quad\hbox{along}\quad\frac{dx}{dt}=u(x(t),t)=K*m,
\label{Lag-inv}
\end{eqnarray}
where again the characteristic velocities depend nonlocally on the quantities that they carry. 
These formulas imply that the signs of the momenta $m$ and $n$ are preserved along characteristics 
during the motion. In particular, if the initial conditions for either of the momenta $m$ and $n$ are everywhere positive, then the sign of that momentum is preserved throughout the subsequent evolution. 

\paragraph{Multi-peakon solutions of CH and CCCH.}

For CH, the velocity is related to the momentum by convolution with
the kernel $K$ in (\ref{kernelK}), which is the Greens function for the Helmholtz operator.
When expressed in terms of the momentum, the multi-peakon solution
of CH for $\omega=0$ is a sum over delta functions,
supported on a set of points moving on the real line. That is, CH for $\omega=0$
has solutions for velocity and momenta given, respectively,  by
\begin{equation}
u(x,t)=\sum_{a=1}^{N}\, p_{a}(t)e^{-|x-q_{a}(t)|}
\quad\hbox{and}\quad
m(x,t)=\sum_{a=1}^{N}\, p_{a}(t)\delta(x-q_{a}(t))
\,.
\label{CHpeakon-m-soln}
\end{equation}
 As shown in \cite{HoMa2004}, the CH peakon solution (\ref{CHpeakon-m-soln})
is geometrically the cotangent-lift momentum map for the left action
of the diffeomorphisms ${\rm Diff}(\mathbb{R})$ on a set of $N$
points on the real line. The same momentum map applies for CCCH, resulting in the 
peakon solutions, with velocities 
\begin{eqnarray}
u(x,t)=\tfrac{1}{2}\sum_{a=1}^{M}m_{a}(t)\, e^{-|x-q_{a}(t)|}\,,
\qquad v(x,t)=\tfrac{1}{2}\sum_{b=1}^{N}n_{b}(t)\, e^{-|x-r_{b}(t)|}
\,,\label{xflow-peakons}
\end{eqnarray}
and momenta supported on delta functions moving along the real line, 
\begin{eqnarray}
m(x,t)=\sum_{a=1}^{M}m_{a}(t)\,\delta(x-q_{a}(t))\,,\qquad n(x,t)=\sum_{b=1}^{N}n_{b}(t)\,\delta(x-r_{b}(t))
\,.\label{xflow-peakons-veloc}
\end{eqnarray}
The dynamics of the CCCH equations and their peakon solutions will be the focus of the remainder of the paper.


\subsection{Main content of the paper}

Section \ref{2-componentCCCH}  studies the two-component solutions of the CCCH equations  (\ref{Xfleq1}) and  (\ref{Xfleq2}). This is done first in a numerical solution of their partial differential equations that establishes their characteristic wave behaviour in the classic dam-break problem for water waves. Then the study of their peakon solutions begins. 

Section \ref{Sec 1} derives the system of cross-coupled CH (CCCH) equations in (\ref{CC-m-eqn}) and  (\ref{CC-n-eqn}) from a variational principle on the direct-product space of $C^0$ vector fields 
$\mathfrak{X}(\mathbb{R}) \times \mathfrak{X}(\mathbb{R})$.

Section \ref{NpeakonSolns-sec} discusses $N$-peakon solutions of the CCCH system.

Section \ref{coupledPpair-sec} studies the simplest possible case, $M=N=1$. This is the peakon couple solution for CCCH.

Section \ref{periodicPeakons-sec} discusses the periodic equilibrium solutions of the CCCH system.

Section \ref{ppbar-solution-sec} considers the anti-symmetric collision of a CCCH peakon couple with its spatial reflection.

Section \ref{othercollisions-sec} presents the results of numerical integrations of the interactions
between the cross-coupled peakon couple and other coherent couple structures for CCCH.

Section \ref{compactons-sec} studies compacton-couple solutions of the CCEP  system (\ref{CCEP-m-eqn}) and (\ref{CCEP-n-eqn}). The dynamics of these compacton couples have some interesting features that depend on the shapes of their profiles. We study triangular and parabolic profiles. 

Section \ref{conclusion-sec} recaps the main results of the paper and lists some of the outstanding 
problems remaining for future investigations of the CCCH and CCEP equations.

\section{Two-component solutions of the CCCH equations} \label{2-componentCCCH}


%
\begin{figure}
\begin{centering}
(a)\includegraphics[clip,width=0.45\textwidth]{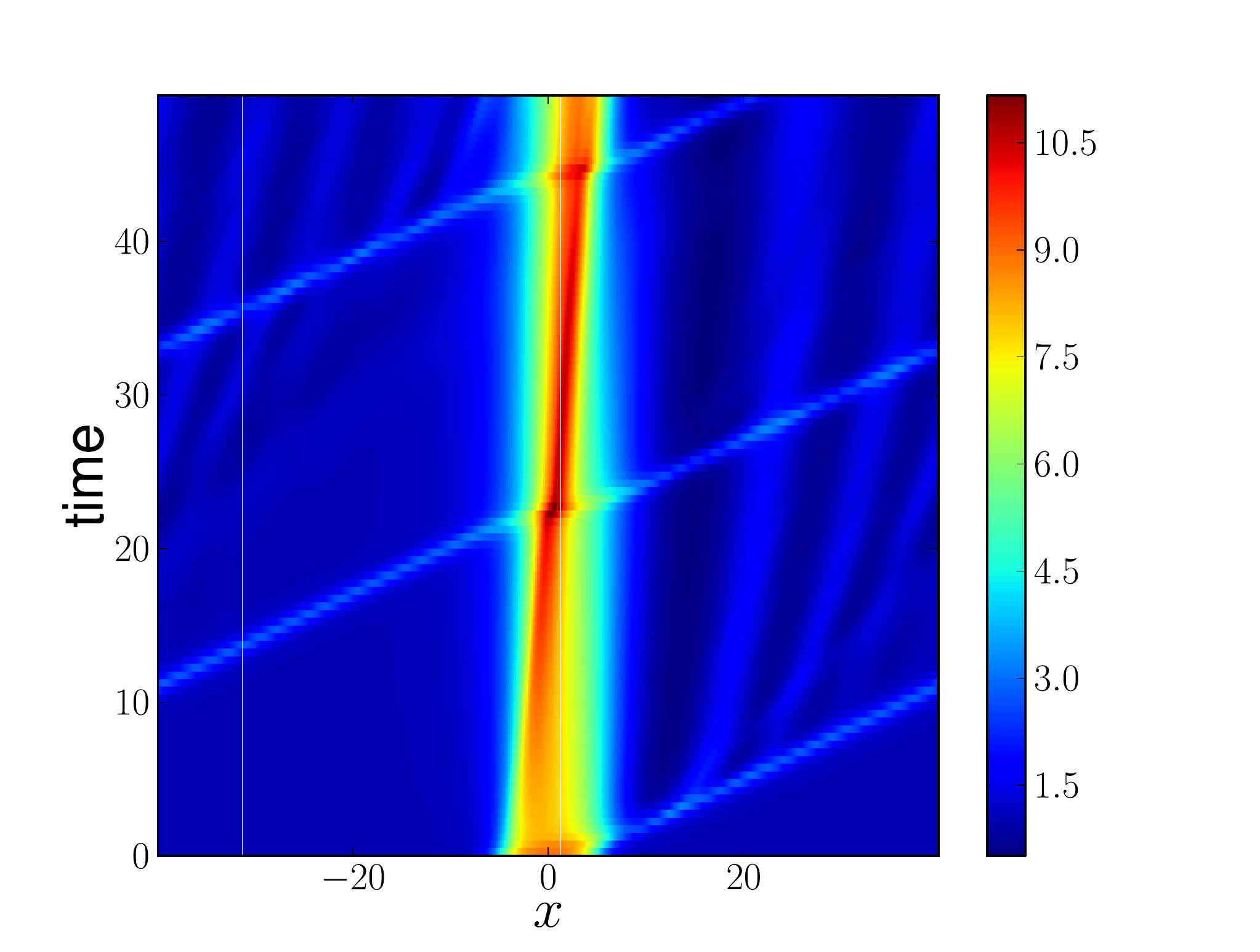} 
(b)\includegraphics[clip,width=0.45\textwidth]{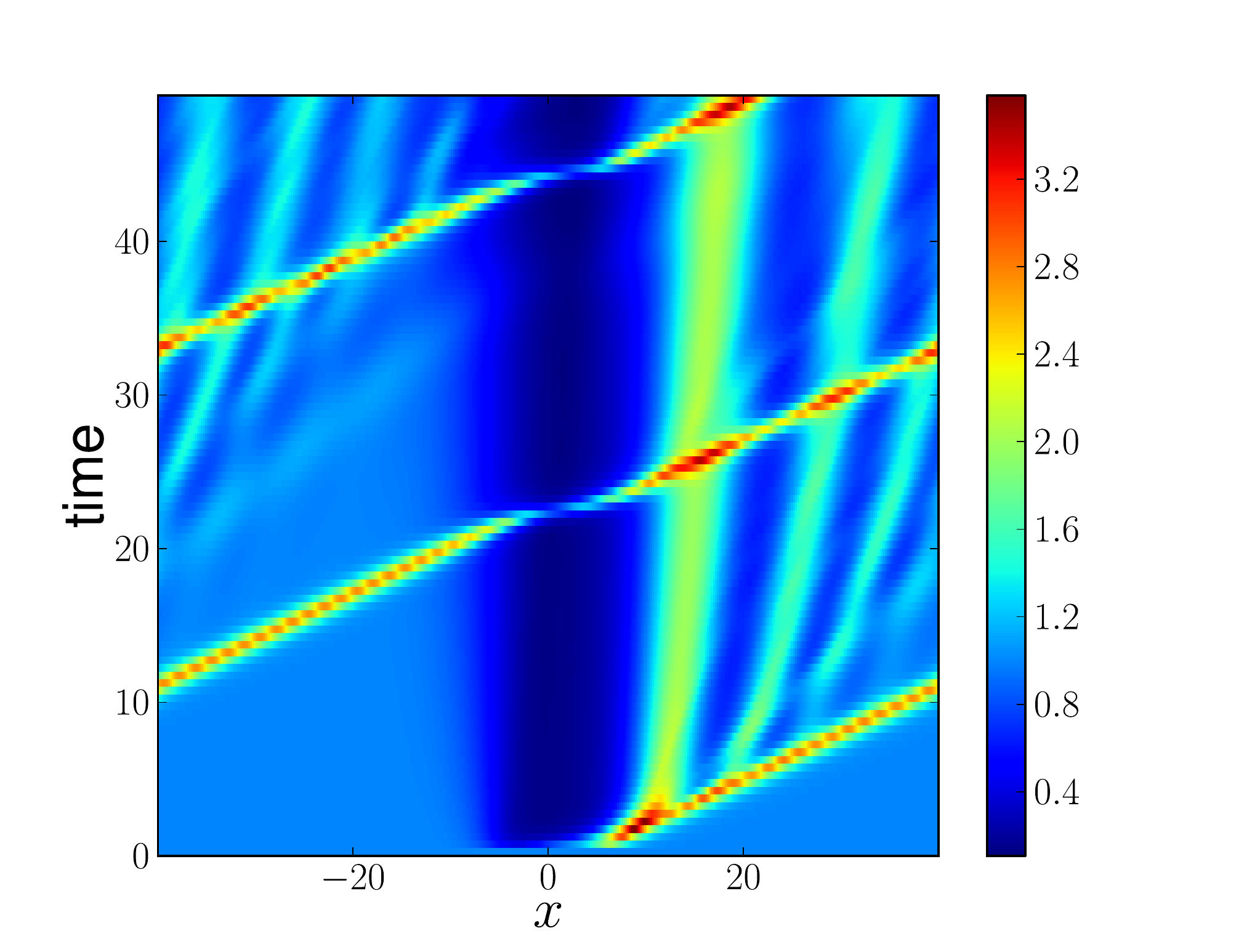} 
\par\end{centering}

\caption{\label{xflow-figs} Spacetime plots of the dam-break solution behaviour
of the velocity variables, $v$ (Panel a) and $u$ (Panel b) of the cross-coupled 
equations (\ref{Xfleq1}, \ref{Xfleq2}) for initial conditions (\ref{eq:dam_conditions}) with tanh-squared
profile and disturbance width 10 units.
The subsequent flow is calculated numerically in a periodic domain of
width 80 units. As in the classical Saint Venant shallow water problem,
wave-like phenomena are created at the moment of release, though the waves propagate
only in a rightward direction due to positive mean velocities.}

\end{figure}

\begin{figure}
(a)\includegraphics[clip,width=0.45\textwidth]{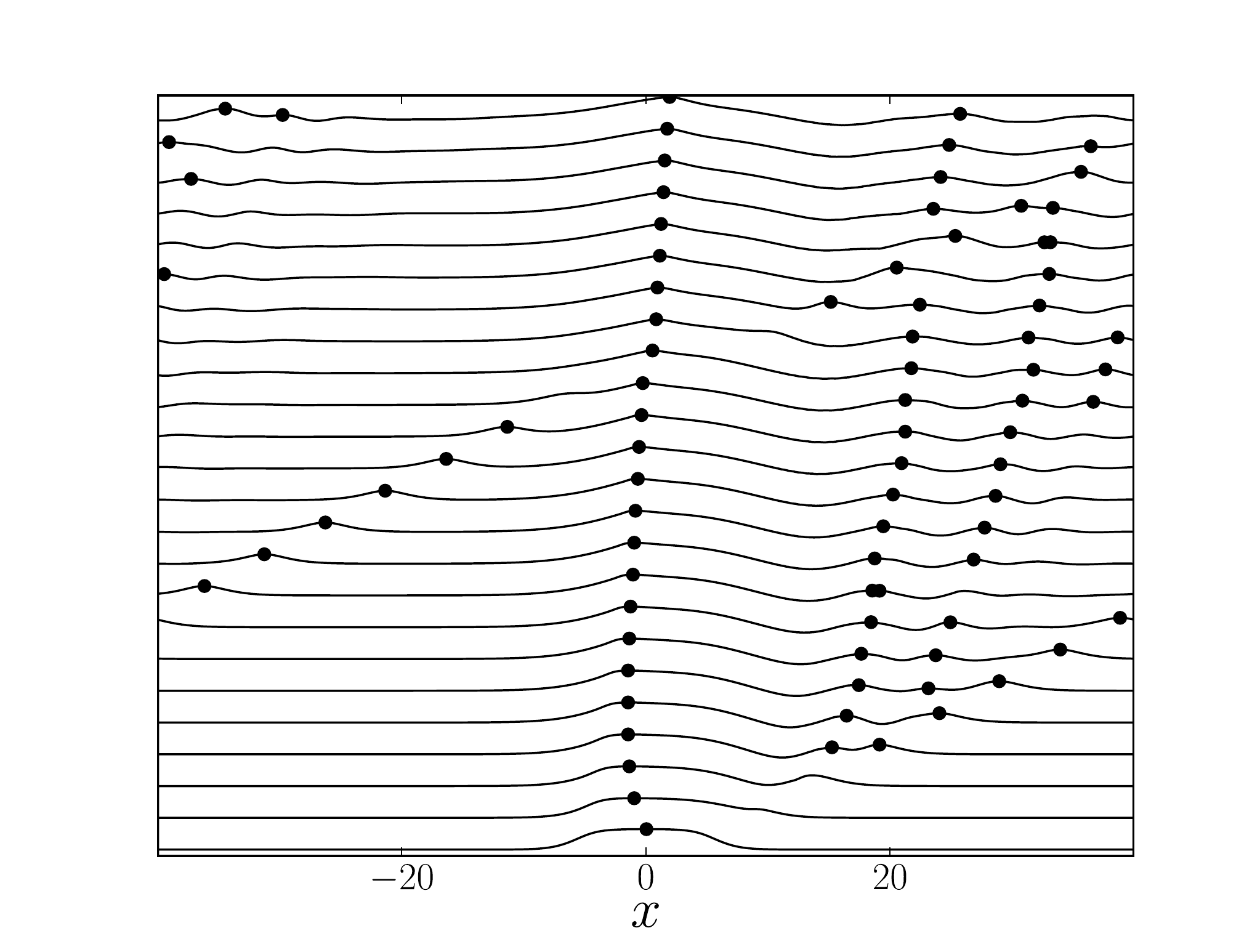} 
(b)\includegraphics[clip,width=0.45\textwidth]{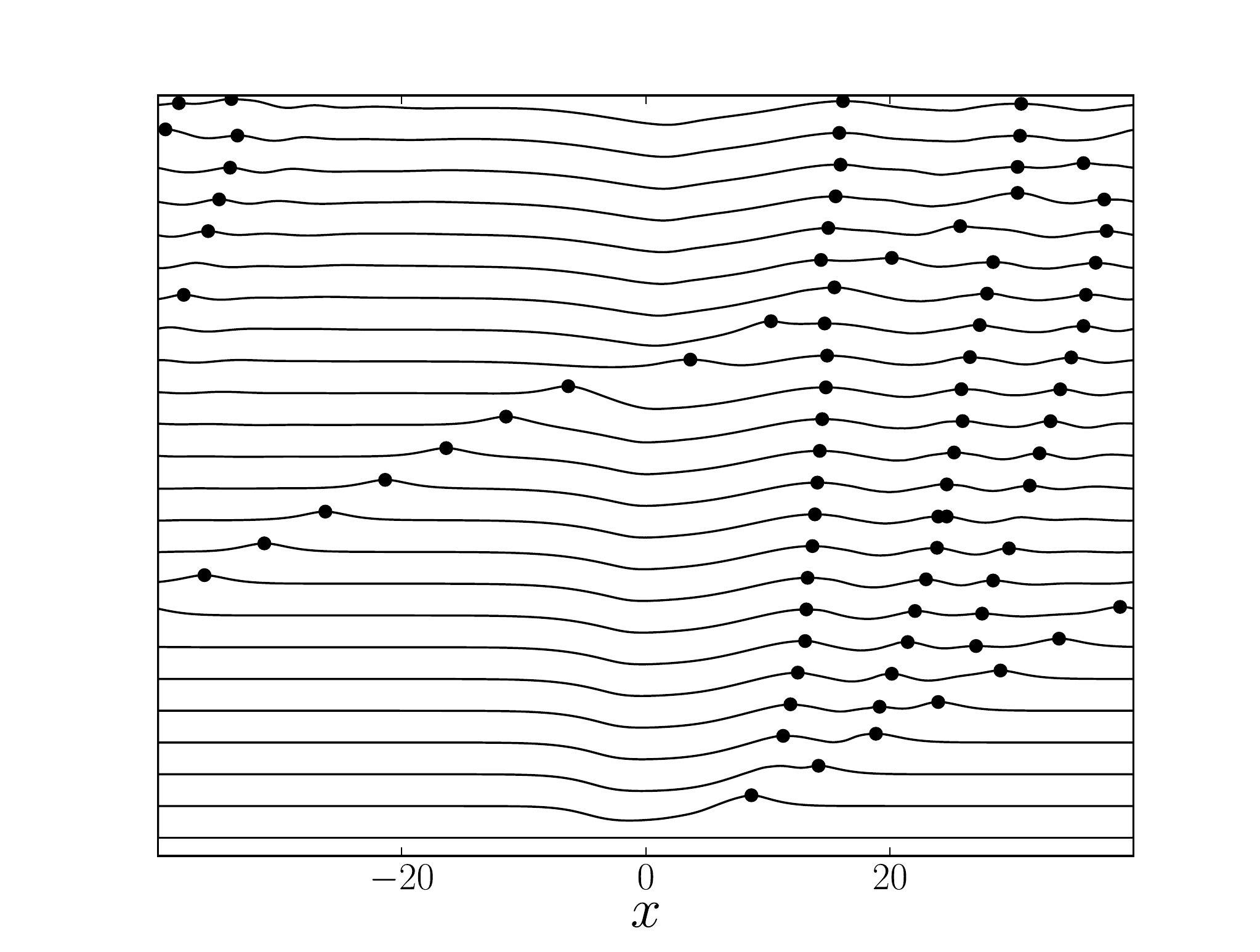} 

\caption{\label{xflow-waterfall-figs} Waterfall plots for the dam-break solution
behaviour of the velocity variables $v$ (Panel a) and $u$ (Panel b) of the
cross-coupled equations (\ref{Xfleq1}, \ref{Xfleq2}) for the same data
as Figure \ref{xflow-figs}. Solutions at intervals of 3 time units are
plotted sequentially up the page and the dots identify locations of the three highest local maxima at each time interval.
\smallskip\\ 
Panel (a) shows the creation of a slow main excitation, a faster excitation and a few slower subsidiary excitations that branch off from it.\smallskip\\ 
Panel (b) shows a fast rightward moving excitation induced by the development of the first fast excitation
in the left panel, and several slow subsidiary excitations that branch off from it. 
\smallskip\\ 
Both of these excitations are coherent structures that seem to retain their identities after the collision
with each other in the middle of the figure.
However, no compelling evidence about the integrability of the CCCH system can be drawn from this data. 
}

\end{figure}

\paragraph{{}`Dam-break' equivalent problem for the cross-coupled equations.}

The dam-break problem involves a body of water of uniform depth, initially
retained behind a barrier, in this case at $x=\pm5$ in a periodic
domain of width 80 units. When the barrier is suddenly removed at
$t=0$, the water flows downward and outward under gravity. The problem
is to find the subsequent flow and determine the shape of the free
surface. This question was addressed in the context of shallow-water
theory, e.g., by Acheson \cite{Ach1990}. The dam-break results for
the cross-coupled system in (\ref{Xfleq1})-(\ref{Xfleq2}) in Figure 
\ref{xflow-figs} show evolution of the variables $v$ (left panel)
and $u$ (right panel), arising from initial conditions in a periodic
domain, in which an initially localized disturbance with tanh-squared
profile in the momentum variable $n$ interacts with an initially constant mean 
flow in the independent cross-coupled velocity field, $u$. That is, 
\begin{eqnarray}
n(x,0) & =&\left[1+\tanh\left(x+5\right)-\tanh\left(x-5\right)\right]^{2},
\label{eq:dam_conditions}\\
m(x,0) & =&u(x,0)=1.
\nonumber 
\end{eqnarray}
 This dam-break initial condition first spawns a rapid leading pulse
in velocity $u$ and then produces a series of slower pulses in $u$,
some of which are of \textit{larger amplitude} than the first pulse.
The rapid first pulse in $u$ then passes again through the periodic
domain, overtaking and colliding with the series of slower but larger
pulses and undergoing a strong interaction indicated by a burst of
amplitude in each collision. The $u$ pulse and the $v$ pulse track
each other, but their variations have opposite phase. That is, a maximum
in the $u$ pulse corresponds to a minimum in the $v$ pulse, and
vice versa. This is an interesting scenario that we will see emerging
when we study the coupled peakon solutions.


\section{Variational derivation of the CCCH and CCEP equations}

\label{Sec 1}

In this section we derive the CCCH and CCEP equations (\ref{CC-m-eqn}) and  (\ref{CC-n-eqn})  from Hamilton's principle, by using the Euler-Poincar\'e theory for symmetry reduction of right-invariant Lagrangians defined on the tangent space of a Lie group.  
We begin by considering motion on the direct product of two copies of the smooth invertible maps with smooth inverses (diffeomeorphisms) acting on the real line ${\rm Diff}(\mathbb{R})\times{\rm Diff}(\mathbb{R})$.  We consider a reduced Lagrangian $l({u},v):\,\mathfrak{X}(\mathbb{R})\times\mathfrak{X}(\mathbb{R})\to\mathbb{R}$
that depends on a pair of smooth vector fields $({u},v)\in\mathfrak{X}(\mathbb{R})=T_e{\rm Diff}(\mathbb{R})/{\rm Diff}(\mathbb{R})$
that are right-invariant under ${\rm Diff}(\mathbb{R})$. Applying the standard
calculation of Hamilton's principle for this right-invariant Lagrangian
yields, see e.g. \cite{HoMaRa1998,HoScSt2009}, \begin{eqnarray*}
\delta\int_{a}^{b}l({u},v)dt & = & \int_{a}^{b}\left\langle \frac{\delta l}{\delta{u}},\delta{u}\right\rangle +\left\langle \frac{\delta l}{\delta{v}},\delta{v}\right\rangle dt\\
 & = & \int_{a}^{b}\left\langle \frac{\delta l}{\delta{u}}\,,\,\frac{d\eta}{dt}-{\rm ad}_{u}{\eta}\right\rangle +\left\langle \frac{\delta l}{\delta{v}}\,,\,\frac{d\xi}{dt}-{\rm ad}_{v}{\xi}\right\rangle dt\\
 & = & -\int_{a}^{b}\,\left\langle \frac{d}{dt}\frac{\delta l}{\delta{u}}+{\rm ad}_{u}^{*}\frac{\delta l}{\delta{u}}\,,\,{\eta}\right\rangle +\left\langle \frac{d}{dt}\frac{\delta l}{\delta{v}}+{\rm ad}_{v}^{*}\frac{\delta l}{\delta{v}}\,,\,{\xi}\right\rangle \,\, dt\,,\end{eqnarray*}
 where one invokes $\eta=0$ and $\xi=0$ for the variations in $\mathfrak{X}(\mathbb{R})$ vanishing at the endpoints in time when integrating by parts and takes the continuum velocity vector fields $({u},v)\in\mathfrak{X}(\mathbb{R})$ to vanish
asymptotically in space. We have also used the formulas \begin{equation}
\delta{u}=\frac{d\eta}{dt}-{\rm ad}_{u}{\eta}\,,\qquad\delta{v}=\frac{d\xi}{dt}-{\rm ad}_{v}{\xi}\,,\quad\hbox{with}\quad\eta,\ \xi\in\mathfrak{X}(\mathbb{R})\,,\label{right-invar-vars}\end{equation}
 for the variation of the right-invariant vectors field $(u,v)$ and have taken the pairing 
\[
\langle\,\cdot\,,\,\cdot\,\rangle:\,\mathfrak{X}(\mathbb{R})\times\mathfrak{X}^{*}(\mathbb{R})\to\mathbb{R}
\]
as the $L^2$ pairing between elements of the Lie algebra $\mathfrak{X}(\mathbb{R})$
and its dual $\mathfrak{X}^{*}(\mathbb{R})$, the space of smooth
1-form densities. 
This pairing allows us to define the ${\rm ad}^{*}$-operation
as 
\begin{equation}
\left\langle {\rm ad}_{v}^{*}\frac{\delta l}{\delta{v}}\,,\,{\xi}\right\rangle =\left\langle \frac{\delta l}{\delta{v}}\,,\,{\rm ad}_{v}{\xi}\right\rangle 
,
\label{ad-star-def}
\end{equation}
 where
 \begin{equation}
{\rm ad}_{v}{\xi}:=-[v,\xi]:=-v\xi_{x}+\xi v_{x}
\label{ad-def}
\end{equation}
 is the adjoint Lie-algebra action $\mathfrak{X}(\mathbb{R})\times\mathfrak{X}(\mathbb{R})\to\mathfrak{X}(\mathbb{R})$,
given by minus the Jacobi-Lie, or commutator bracket of smooth vector
fields. In our case, $\langle\cdot\,,\,\cdot\rangle$ is the $L^{2}$
pairing between vector fields $\mathfrak{X}(\mathbb{R})$ and 1-form
densities $\mathfrak{X}^{*}(\mathbb{R})$, written as 
\[
\left\langle \frac{\delta l}{\delta{u}},\delta{u}\right\rangle =\int\delta{u}\,\frac{\delta l}{\delta{u}}\, dx
,
\]
 in which the integral is taken over the infinite real line.

Substituting the reduced Legendre transformation
$u \mapsto \mu = \delta \ell / \delta u$ and $v \mapsto \nu = \delta \ell / \delta v$ 
produces the following pair of Euler-Poincar\'e equations governing coadjoint motion for $(\mu, \nu)\in\mathfrak{X}^*(\mathbb{R})\times\mathfrak{X}^*(\mathbb{R})$,
\begin{equation}
\partial_{t} \mu +{\rm ad}_{u}^{*} \mu =0
\,,\quad
\partial_{t} \nu +{\rm ad}_{v}^{*} \nu =0
\quad\hbox{for}\quad
\mu:=\delta l/\delta{u}
\,,\quad
\nu:=\delta l/\delta{v}
\,.
\label{EP-eqns}
\end{equation}
 We may evaluate the ${\rm ad}^{*}$-expressions from their definitions,
so that 
\[
\left\langle {\rm ad}_{}^{*}\mu\,,\,{\xi}\right\rangle 
=\left\langle \mu \,,\,{\rm ad}_{u}{\xi}\right\rangle 
=\left\langle \mu\,,\,-u\xi_{x}+\xi u_{x}\right\rangle 
=\left\langle (\mu u)_{x}+\mu u_{x},\,\xi\right\rangle 
\,,
\]
and similarly for $(\mu,u)\to(\nu,v)$.
These expressions yield the Euler-Poincar\'e equations on the real line for an arbitrary right-invariant Lagrangian.

If the reduced Legendre transformation $(u,v) \mapsto (\mu,\nu)$ 
is invertible, then the Euler-Poincar\'e equations (\ref{EP-eqns}) are equivalent
to the (right) Lie-Poisson Hamiltonian equations:
\begin{equation}
\label{lpequations} 
\dot \mu  =  - {\rm ad} ^{* }_{\delta h/\delta \mu} \mu
\,,\quad
\dot \nu  =  - {\rm ad} ^{* }_{\delta h/\delta \nu} \nu
\,, 
\end{equation} 
where the reduced Hamiltonian $h$ and its variational derivatives are obtained from the reduced Legendre transformation by
\[
h (\mu,\nu) = \left\langle \mu, u\right\rangle + \left\langle \nu, v\right\rangle - l({u},v).
\]
These equations are equivalent (via Lie-Poisson reduction and
reconstruction) to Hamilton's equations on $T ^{\ast} {\rm Diff} \times T^\ast {\rm Diff} $
relative to the Hamiltonian $H :T ^{\ast} {\rm Diff} \times T^\ast {\rm Diff}  \rightarrow
\mathbb{R}$, obtained by right translating $h$ from the identity
element to other points via the right (diagonal) action of ${\rm Diff}$ on $T ^{\ast} {\rm Diff} \times T^\ast {\rm Diff} $.
The Lie-Poisson equations may be written in the Poisson bracket
form
\begin{equation} \label{lppoisson}
\dot{F}(\mu,\nu ) = \left\{ F, h \right\},
\end{equation}
where $F: \mathfrak{X}^*\times\mathfrak{X}^* \rightarrow \mathbb{R}$ is
an arbitrary smooth function and the bracket is the
(right) Lie-Poisson bracket given by
\begin{equation}
\label{lpb} \{F, G\}(\mu,\nu )  
=  \left\langle \mu , \left[ \frac{ \delta F}{\delta  \mu},
\frac{\delta  G}{\delta \mu } \right] \right\rangle 
+
\left\langle \nu , \left[ \frac{ \delta F}{\delta  \nu},
\frac{\delta  G}{\delta \nu } \right] \right\rangle . 
\end{equation} 

In the cases we shall consider, the Lagrangian $l(u,v)$ will be bilinear, and its variations will cross-couple the momentum equations, as follows. An arbitrary norm on the tangent space of vector fields,
 $\left\Vert u\right\Vert =\left\langle\,u,\mathcal{G} u\right\rangle,$
with $\mathcal{G}$ a positive symmetric operator on $u$,
induces cross-coupled equations through the application of the Euler-Poincar\'e framework
to the new Lagrangian
$
l_{\mathcal{G}}\left(u,v\right) = \left\langle\,u,\mathcal{G} v\right\rangle 
=\left\langle\,\mathcal{G} u,v\right\rangle .
$
 In particular, the momenta in the Legendre transformation are given by,
\[
\mu=\frac{\delta l_{\mathcal{G}}}{\delta u}=\mathcal{G}v
\quad\hbox{and}\quad
\nu=\frac{\delta l_{\mathcal{G}}}{\delta v}=\mathcal{G}u.
\]

The corresponding Hamiltonian is given by
\[
h_\mathcal{G}(\mu,\nu)= \left\langle\,K_\mathcal{G}*\mu,\nu\right\rangle
= \left\langle\,K_\mathcal{G}*\nu,\mu\right\rangle
\]
by symmetry of the kernel $K_\mathcal{G}$. 
Consequently, the associated velocities are obtained from the convolutions 
\[
\frac{\delta h_{\mathcal{G}}}{\delta \mu} = K_\mathcal{G}*\nu = u
\quad\hbox{and}\quad
\frac{\delta h_{\mathcal{G}}}{\delta \nu} = K_\mathcal{G}*\mu = v
\]
in which the kernel $K_\mathcal{G}$ for the system is the Greens function for the invertible operator $\mathcal{G}$. The resulting Euler-Poincar\'e or Lie-Poisson equations may both be expressed as
\begin{equation}
\partial_{t} \mathcal{G}v +{\rm ad}_{u}^{*} \mathcal{G}v =0
\,,\quad
\partial_{t} \mathcal{G}u +{\rm ad}_{v}^{*} \mathcal{G}u =0
\,.
\label{CCEP-eqns}
\end{equation}
These are the equations of the CCEP system (\ref{CCEP-m-eqn}) and (\ref{CCEP-n-eqn}) for a general choice of the operator $\mathcal{G}$ and its Greens function kernel $K_\mathcal{G}$. They will be the CCCH equations (\ref{CC-m-eqn}) and  (\ref{CC-n-eqn}) with $\omega=0$, for the choice of the Helmholtz operator, $\mathcal{G}=1-\partial_x^2$.

Let us consider a few choices of the bilinear reduced Lagrangian $\ell({u},v)$
whose Euler-Poincar\'e equations (\ref{CCEP-eqns}) support peakon solutions on the real
line: 
\begin{enumerate}
\item When the reduced Lagrangian depends on only one vector field as $\ell({u})=\frac{1}{2}\|{u}\|_{H^{1}}^{2}$,
then the dispersionless CH equation, \[
\partial_{t}m+um_{x}+2mu_{x}=0\,,\quad\hbox{with}\quad m=u-u_{xx},\]
 emerges as the dynamics of geodesic motion on the diffeomorphisms
with respect to the $H^{1}$ norm $\|{u}\|_{H^{1}}=\int u^{2}+u_{x}^{2}\, dx$. 
\item When the reduced Lagrangian is taken as $\ell({u},v)=\int uv+u_{x}v_{x}\, dx$,
we find the following pair of coupled Euler-Poincar\'e equations, in this case for the symmetric positive operator $\mathcal{G}=1-\partial_x^2$, the Helmholtz operator,
\begin{equation}
\partial_{t}m+2v_{x}m+vm_{x}=0\,,
\qquad\partial_{t}n+2u_{x}n+un_{x}=0\,,
\label{cross-flow-eqns}
\end{equation}
 with $m=\delta l/\delta{v}=u-u_{xx}=\mathcal{G}u$ and $n=\delta l/\delta{u}=v-v_{xx}=\mathcal{G}v$. Because the momentum density
($m$) for one velocity vector field ($v$) is Lie-dragged by the
vector field ($u$) for the other momentum density ($n$), we call
this system the {\bfi cross-coupled equations}. When $v=u$, this
coupled system restricts to the dispersionless CH equation. (Dispersion
may be introduced by adding constant shifts to variables $m=u-u_{xx}+\omega_{1}$ and $n=v-v_{xx}+\omega_{2}$.) 
\item When $(u,v)\in\mathbb{C}$ and $v=\bar{u}$, the reduced Lagrangian
becomes the complex $H^{1}$ norm $\ell({u})=\int(|u|^{2}+|u_{x}|^{2})\, dx$
and we may interpret the coupled system of cross-coupled equations (\ref{cross-flow-eqns})
as the complex version of the dispersionless CH equation. 
\item The system of cross-coupled equations (\ref{cross-flow-eqns})
may be written in Lie-derivative form as
\begin{equation}
(\partial_{t}+\mathcal{L}_{v})\big(m\, dx^{2}\big)=0\quad\hbox{and}\quad(\partial_{t}+\mathcal{L}_{u})\big(n\, dx^{2}\big)=0\,,\label{cross-flow-eqns-Lie}\end{equation}
 with two characteristic velocities, $u=K*m$ and $v=K*n$, in which
the kernel $K$ is the Green's function for the Helmholtz operator,
defined in (\ref{kernelK}). The cross-coupled equations may also be
written equivalently in a form similar to equation (\ref{Lag-inv}), as follows, 
\begin{eqnarray}
\frac{d}{dt}\big(m\, dx^{2}\big) & = & 0\quad\hbox{along}\quad\frac{d{x}}{dt}={v({x},t)}=K*n\,,\label{cross-flow-eqn1-Lag}\\
\frac{d}{dt}\big(n\, dx^{2}\big) & = & 0\quad\hbox{along}\quad\frac{d{x}}{dt}={u({x},t)}=K*m\,.\label{cross-flow-eqn2-Lag}\end{eqnarray}
These equivalent forms of the cross-coupled equations imply they admit 
particle-like delta function solutions for both $m$ and $n$, as discussed in the next section.

\end{enumerate}

The remainder of the paper is devoted to studying the solutions of the CCCH system (\ref{cross-flow-eqns}) and the more general CCEP system (\ref{CCEP-eqns}).

\section{Peakon solutions of the cross-coupled equations} \label{NpeakonSolns-sec}
We have already asserted that the solutions of the CCCH system (\ref{cross-flow-eqns}) may be expressed in  the form (\ref{xflow-peakons-veloc}) for the two momenta and (\ref{xflow-peakons}) for their corresponding velocities. 
The delta-function representation of these momentum solutions is also the cotangent lift momentum map for the direct product ${\rm Diff}(\mathbb{R})\times{\rm Diff}(\mathbb{R})$ of diffeomorphisms acting on the real line $\mathbb{R}$ by composition from the left \cite{HoMa2004}.

By general principles for momentum maps and in parallel with the peakon dynamics for the CH equation, the $2M+2N$ variables $(q_{a},m_{a})$, $a=1,\dots,M$, and $(r_{b},n_{b})$,
$b=1,\dots,N$, in the peakon solutions for position (\ref{xflow-peakons}) and momentum (\ref{xflow-peakons-veloc}) of the CCCH equations (\ref{Xfleq1}, \ref{Xfleq2}) are governed by Hamilton's canonical equations
for the Hamiltonian function, 
\begin{eqnarray}
H=\tfrac{1}{2}\sum_{a,b=1}^{M,N}m_{a}(t)n_{b}(t)e^{-|q_{a}(t)-r_{b}(t)|}
\,.
\end{eqnarray}
These  canonical equations comprise the evolution equations,  
\begin{eqnarray}
\dot{q}_{a}(t) & = & \frac{\partial H}{\partial m_{a}}=\tfrac{1}{2}\sum_{b=1}^{N}n_{b}(t)e^{-|q_{a}(t)-r_{b}(t)|}=v(q_{a}(t),t)\,,\label{qa}\\
\dot{r}_{b}(t) & = & \frac{\partial H}{\partial n_{b}}=\tfrac{1}{2}\sum_{a=1}^{M}m_{a}(t)e^{-|q_{a}(t)-r_{b}(t)|}=u(r_{b}(t),t)\,,\label{rb}\end{eqnarray}
for the positions of the peakons, and \begin{eqnarray}
\dot{m}_{a}(t) & = & -\,\frac{\partial H}{\partial q_{a}}=\tfrac{1}{2}m_{a}\sum_{b=1}^{N}n_{b}\,{\rm sgn}\,(q_{a}-r_{b})e^{-|q_{a}(t)-r_{b}(t)|}=-\, m_{a}\,\frac{\partial v}{\partial x}\Big|_{x=q_{a}}\,,\label{ma}\\
\dot{n}_{b}(t) & = & -\,\frac{\partial H}{\partial r_{b}}=-\tfrac{1}{2}n_{b}\sum_{a=1}^{M}m_{a}\,{\rm sgn}\,(q_{a}-r_{b})e^{-|q_{a}(t)-r_{b}(t)|}=-\, n_{b}\,\frac{\partial u}{\partial x}\Big|_{x=r_{b}}\,,\label{nb}\end{eqnarray}
for their canonical momenta. In the case originally envisioned that the solutions would be complex, for $v=\bar{u}\in\mathbb{C}$ we have $N=M$ and $n_{a}=\bar{m}_{a}$.

Conserved quantities for these equations include the energy $H$ and the total momentum
$\sum_{a}(m_{a}+n_{a})$. The solutions of these peakon equations will be studied in
more detail in the sections that follow.

\section{The Peakon Dipole Solution} \label{coupledPpair-sec}

In this section, we consider the peakon solutions of the CCCH system in the simplest possible case, $M=N=1$. Dropping
superfluous indices for this case, we introduce new canonical position variables
$X=\left(q+r\right)/2$, $Y=q-r$, respectively mean position of the
peaks and their separation distance. The evolution equations in terms
of the new variables are\[
\dot{X}=\frac{\left(m+n\right)}{4}e^{-|Y|},\]
 \[
\dot{Y}=\frac{\left(n-m\right)}{2}e^{-|Y|}.\]
Thus we can define the behaviour of the exponential function of the
absolute separation of the peaks, \begin{equation}
\frac{d}{d t}e^{|Y|}=\sgn\left(Y\right)\frac{\left(n-m\right)}{2},\label{eq:del_dt}\end{equation}

where $\sgn\left(x\right)$ denotes the signum function\[
\sgn\left(x\right)
=
\theta\left(x\right)-\theta\left(-x\right)
=
\left\{\begin{array}{cc}
1 & \hbox{for } x>0,\\
-1 & \hbox{for } x<0.
\end{array}
\right.
\]
Meanwhile
\begin{equation}
\dot{m}=-\dot{n}
=
\sgn\left(Y\right)\frac{mn}{2}e^{-|Y|}=\sgn\left(Y\right)E,
\label{eq:m_eq}
\end{equation}
where $\mbox{E=\ensuremath{\left.H\right|_{t=0}}}$is the (constant)
value of the Hamiltonian, that is to say the total energy of the coupled
pair. Differentiating (\ref{eq:del_dt}) again with respect to time
gives 
\[
\frac{d^{2}}{d t^{2}}\left(e^{|Y|}\right)
=
-\sgn^{2}\left(Y\right)E+\delta\left(Y\right)\left(n-m\right)^{2}
.
\]
On integrating for a particular signature of $\left.Y\right|_{t=0}=Y_{0}\neq0$,
\begin{equation}
e^{|Y|}=-\,\frac{m_{0}n_{0}e^{-|Y_{0}|}t^{2}}{2}+\sgn\left(Y_{0}\right)
\frac{\left(n_{0}-m_{0}\right)}{2}t+e^{|Y_{0}|}
,\label{eq:el}
\end{equation}
where
\[
m_{0}=\left.m\right|_{t=0},\quad n_{0}=\left.n\right|_{t=0}\quad
Y_{0}=\left.Y\right|_{t=0}.
\]
 If $m_{0}$ and $n_{0}$ have the same signature, then eventually
we will have $|Y|=0$, regardless of the value of $|Y_{0}|$. If $m_{0}$
and $n_{0}$ are of different signature, then provided 
$\sgn\left(Y_{0}\right)\left(n_{0}-m_{0}\right)<0$
the same fate will occur. If the signatures differ and 
$\sgn\left(Y_{0}\right)\left(n_{0}-m_{0}\right)>0$
then the particles immediately separate and continue to do so subject
to the logarithmic rate of equation (\ref{eq:el}). In this case the
amplitudes of both particles grow linearly according to equation (\ref{eq:m_eq}).

At the point of collision when $|Y|=0$ we require both $H=\tfrac{1}{2}mn$ and
$\mathcal{M}=m+n$ to be conserved. Thus the absolute value of $m-n$ is also conserved.
If the particles are assumed to `tunnel' through each other then the
momenta are conserved and the signature of the whole problem in $Y$
switches, with the forcing on the amplitudes of the momenta reversed.
Equivalently the particles can be assumed to collide elastically and
to exchange both the amplitude and type of their momenta. In this
interpretation, the signature of the forcings on the particles
is set by the initial configuration.

When $m_{0}$ and $n_{0}$ share the same signature the half period
of their `waltzing' motion can be found
by setting $Y_{0}=0$ and looking for when $e^{|Y|}$ attains unity,
namely $t=2(n_{0}-m_{0})/{n_{0}m_{0}}.$ It will be noted that
at this time
\[
\left.m\right|_{t=2\frac{m_{0}-n_{0}}{m_{0}n_{0}}}
= m_{0}+m_{0}n_{0}\left(\frac{m_{0}-n_{0}}{m_{0}n_{0}}\right)=n_{0},
\]
and similarly
\[
\left.n\right|_{t=2\frac{m_{0}-n_{0}}{m_{0}n_{0}}}=m_{0},
\]
 so that the two types of peakons do indeed exchange momentum amplitudes
over a half cycle. The maximum separation of the particles occurs
when the momenta are equal,
\[
t=\frac{n_{0}-m_{0}}{n_{0}m_{0}},\quad m=n=\frac{n_{0}+m_{0}}{2},\]
 at which point \[
|Y|=\ln\left(1+\frac{\left(n_{0}-m_{0}\right)^{2}}{4m_{0}n_{0}}\right).
\]

For two particles of arbitrary initial separation $Y_{0}$, the full
evolution of the `particles' with respect to each other is described
in terms of the conserved energy,
\[
E=\frac{1}{2}m_{0}n_{0}e^{-\left|Y_{0}\right|}
,
\]
and total momentum\[
\mathcal{M}=m_{0}+n_{0}\]
in terms of the time variable, $t$, by equations 
\[
m\left(t\right)
=
\left\{
\begin{array}{cc}
m_{0}+\overline{\sgn}\left(Y_{0}\right)Et & \hbox{for } t\leq T_{0}\\
m_{0}+\overline{\sgn}\left(Y_{0}\right)E\left(T_{0}-\left(t-T_{2i}\right)\right) & \hbox{for } T_{2i}\leq t\leq T_{2i+1}
\\
n_{0}-\overline{\sgn}\left(Y_{0}\right)E\left(T_{0}+\left(t-T_{2i+1}\right)\right) & \hbox{for } T_{2i+1}\leq t\leq T_{2i}
\end{array}
\right.
\]
 \[
n\left(t\right)
=
\left\{
\begin{array}{cc}
n_{0}-\overline{\sgn}\left(Y_{0}\right)Et &  \hbox{for } t\leq T_{0},\\
n_{0}-\overline{\sgn}\left(Y_{0}\right)E\left(T_{0}+\left(t-T_{2i}\right)\right) &  \hbox{for } T_{2i}\leq t\leq T_{2i+1},\\
m_{0}+\overline{\sgn}\left(Y_{0}\right)E\left(T_{0}-\left(t-T_{2i+1}\right)\right) & \hbox{for }  T_{2i+1}\leq t\leq T_{2i},\end{array}
\right.
\]
 \[
|l\left(t\right)|
=
\left\{
\begin{array}{cc}
\ln\left(-\tfrac{1}{2}Et^{2}+\tfrac{1}{2}\overline{\sgn}\left(Y_{0}\right)t\left(n_{0}-m_{0}\right)+e^{|Y_{0}|}\right) &  \hbox{for } t\leq T_{0},\\
\ln\left(1-\tfrac{1}{2}E\left(t-T_{i}\right)^{2}+\frac{1}{2}\left(t-T_{i}\right)\sqrt{\mathcal{M}^{2}-8E}\right) & 
 \hbox{for } T_{i}\leq t\leq T_{i+1},\end{array}
\right.
\]

 \[
\sgn\left(Y\left(t\right)\right)
=
\left\{
\begin{array}{cc}
\overline{\sgn}\left(Y_{0}\right), & \hbox{for }  t<T_{0},\\
\overline{\sgn}\left(Y_{0}\right)\left(-1\right)^{i}, & \hbox{for }  T_{i}<t\leq T_{i+1},\end{array}
\right.
\]
 where we have defined a modified signum function, 
 \[
\overline{\sgn}\left(Y_{0}\right)
=
\left\{
\begin{array}{cc}
1 & \hbox{for }  Y_{0}>0,\\
\sgn\left(n_{0}-m_{0}\right) & \hbox{for }  Y_{0}=0,\\
-1 & \hbox{for }  Y_{0}<0.\end{array}
\right.
\]
 Here $T_{0}$ is the time until the initial collision, where if $E>0$ or
$\sgn\left(Y_{0}\right)\left(n_{0}-m_{0}\right)\leq0$, this time is given by
\[
T_{0}=\frac{\sgn\left(Y_{0}\right)\left(n_{0}-m_{0}\right)+\sqrt{\mathcal{M}-8E}}{2E},\]
 and $T_{0}=\infty$ otherwise. The subsequent $T_{i}$'s in the peakon-peakon
and antipeakon-antipeakon cases are the times of the subsequent collisions,
\[
T_{i+1}-T_{i}=\frac{\sqrt{\mathcal{M}^{2}-8E}}{E}.\]

%
\begin{figure}[ht]
\begin{centering}
(a)\includegraphics[clip,width=0.45\textwidth]{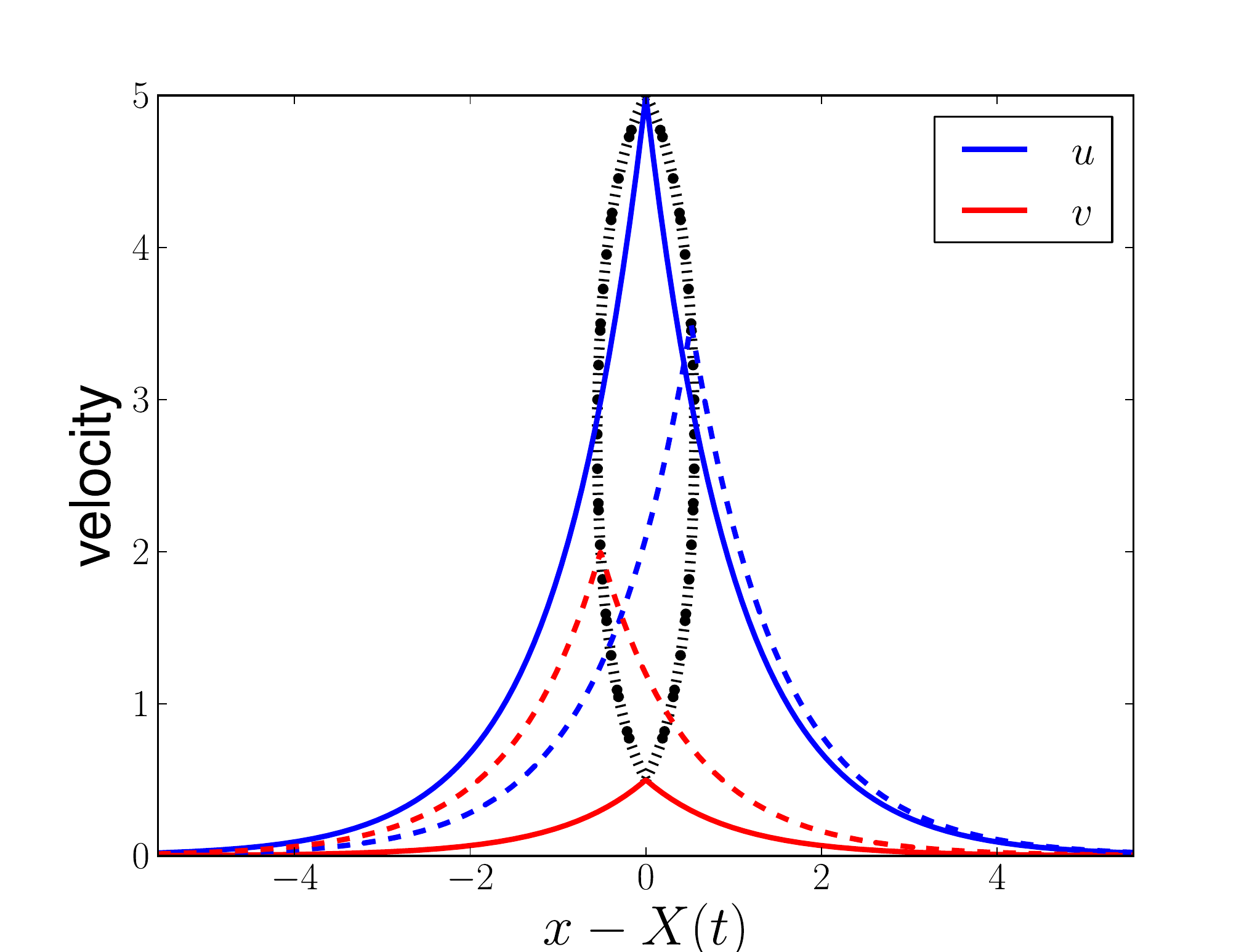}
(b)\includegraphics[clip,width=0.45\textwidth]{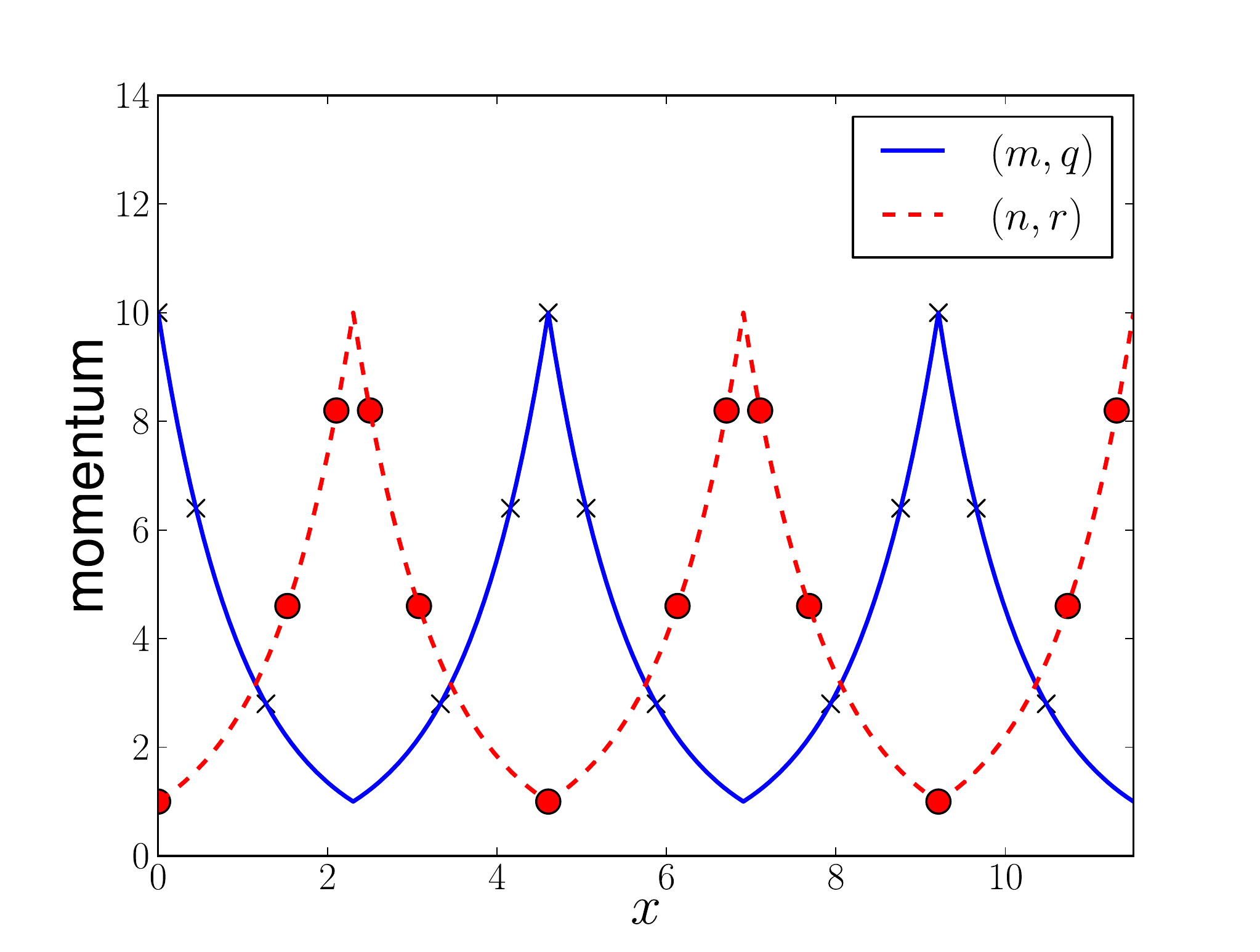} 
\par\end{centering}

\caption{\label{peakon_orbits} Evolution of the velocity (Panel a) and momentum (Panel b) solutions of a coupled peakon pair.\smallskip\\ 
 Panel (a) shows the profile of velocity fields 
of a peakon-peakon couple with initial parameters $m_{0}=10$, $n_{0}=1$,
$Y_{0}=0$ (solid lines). The dotted path indicates the evolution 
of the two peaks in the frame travelling at the particles' mean
velocity, $\dot{X}=\left(\dot{q}+\dot{r}\right)/2$. For these initial
conditions, the total period of one orbit of the cycle is $T=3.6$.
Also shown are the velocity fields at subsequent time intervals $t=0.45+1.8k$
for $k\in\mathbb{N}$ (dashed lines). \smallskip\\ 
Panel (b) shows locus of the magnitude
of momentum versus position for the two particles. The circles and
crosses indicate the respective positions of each particle in phase
space at time intervals of 1.44 units. The peakons are initially
superposed at $x=0$, so that the leftmost circle and cross represent
the same instant, and similarly for each couple continuing rightwards.\\
Animations showing the time evolution of both these images are available
as supplementary material with the online copy of this paper.}
\end{figure}

\begin{figure}[t]
\begin{centering}
(a)\includegraphics[clip,width=0.45\textwidth]{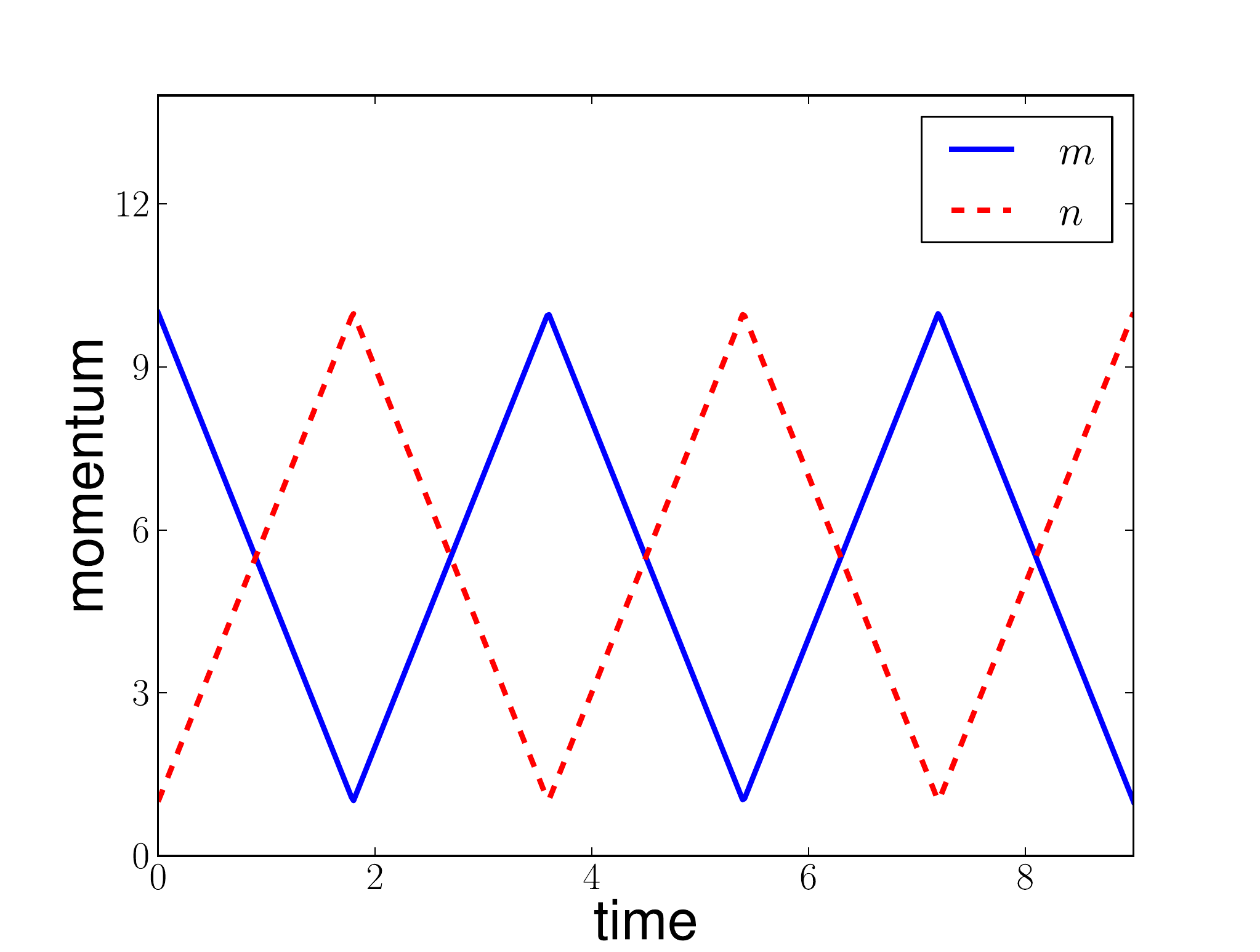}
(b)\includegraphics[clip,width=0.45\textwidth]{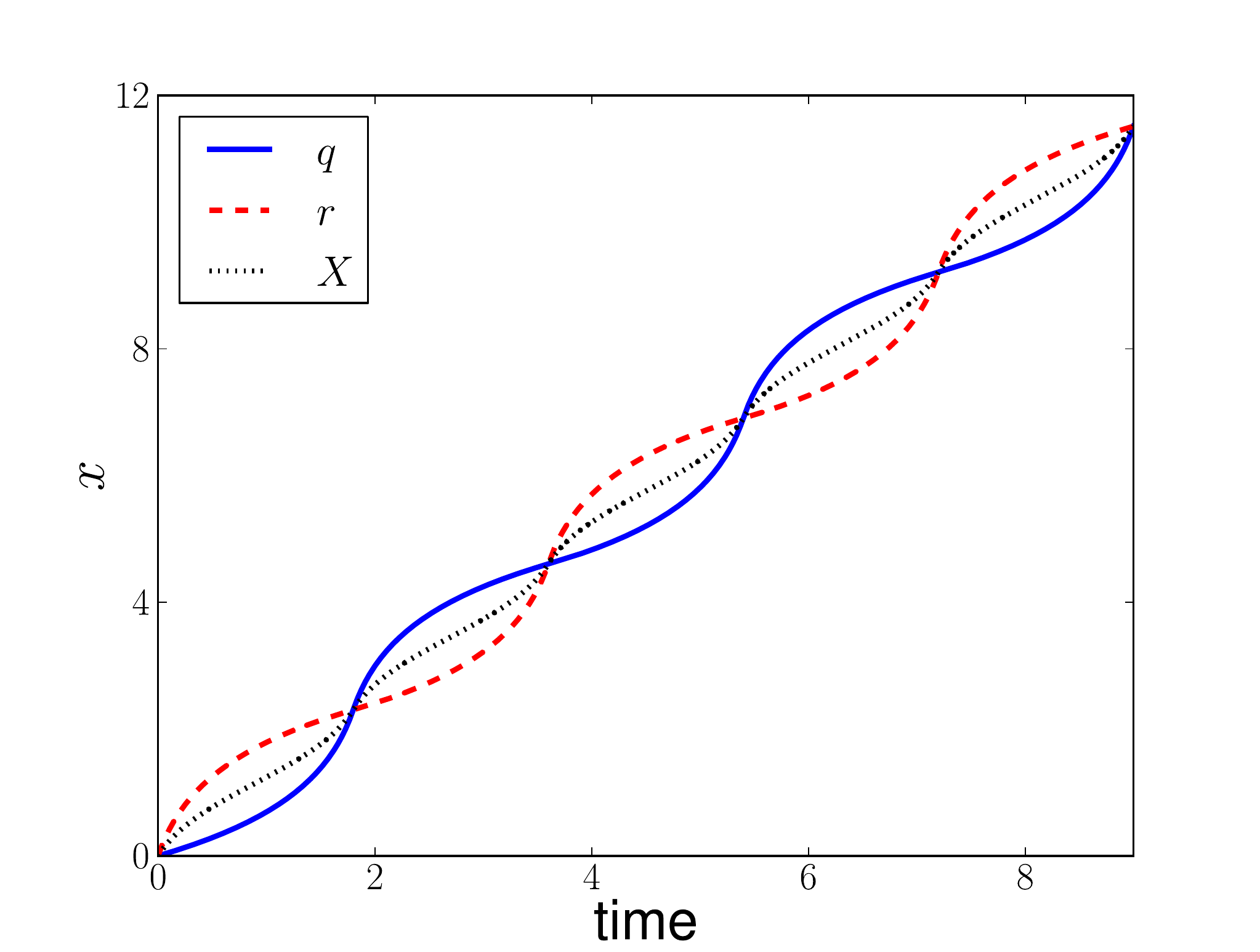} 
\par\end{centering}
\caption{
\label{moms}
Plots of momentum (Panel a) and position (Panel b)
versus time for the peakon-peakon couple with initial conditions
$m_{0}=10$, $n_{0}=1$, $q_{0}=r_{0}=0$. \smallskip\\ 
Panel (a) illustrates that, as per
equation (\ref{eq:m_eq}) the growth and decay of momenta are continuous
and piecewise linear with respect to time in the interval between
collisions, while the total momentum, $\mathcal{M}=m+n$, is conserved through
collisions. Also shown is the discontinuity (and symmetry) in the
first derivative of the momentum at the points of collision, which
here occur every 1.8 time units.\smallskip\\ 
Panel (b) illustrates the {}``waltzing'' nature of the collision, with the identity of the
leading particle alternating periodically, with the locus of the position
of one particle obtainable from the other under a phase shift. Also
plotted is the mean position $X=\left(q+r\right)/2$, showing the
nonlinear relations in the propagation speed of the structure.}
\end{figure}
\begin{figure}[t]
\begin{centering}
(a)\includegraphics[clip,width=0.45\textwidth]{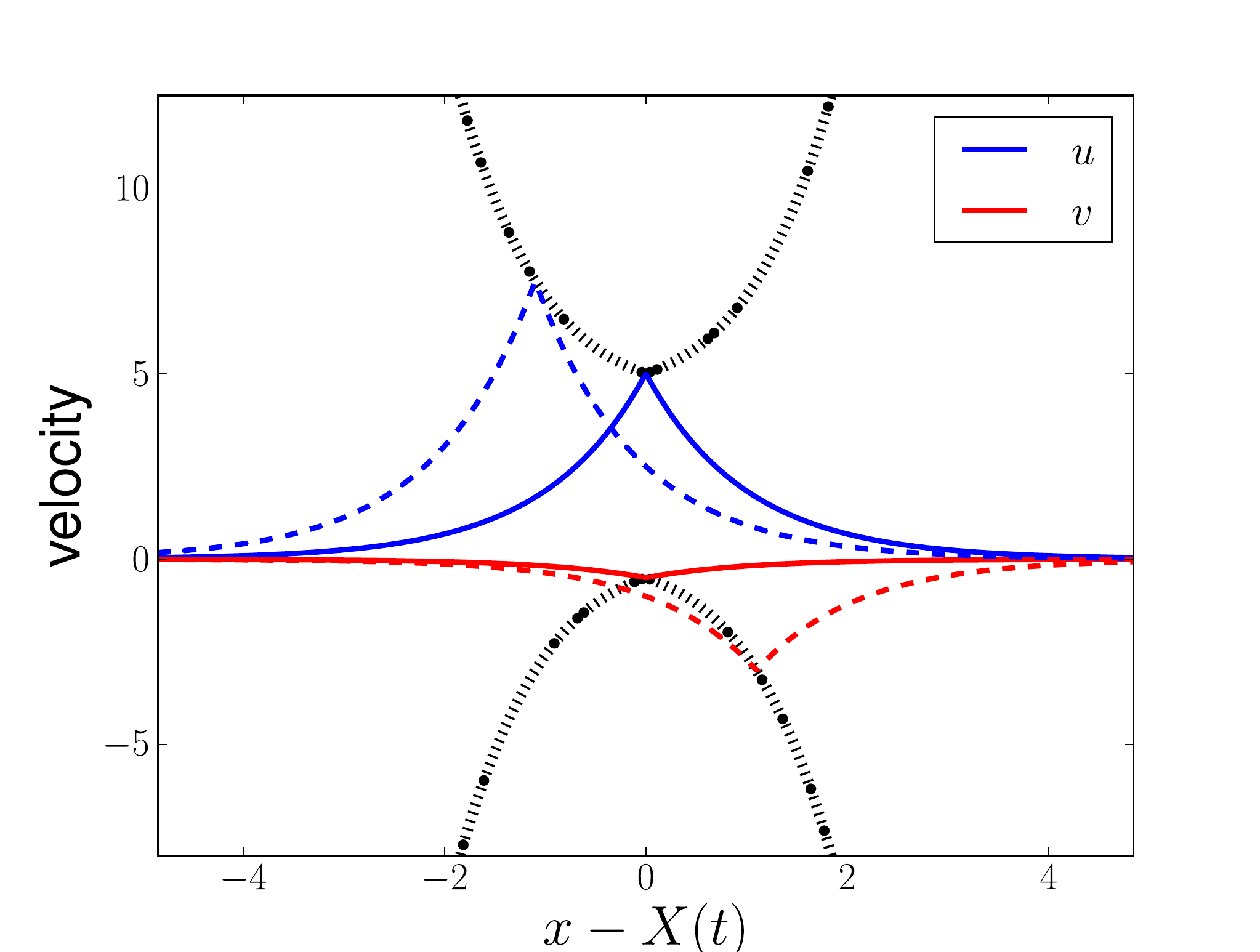}
(b)\includegraphics[clip,width=0.45\textwidth]{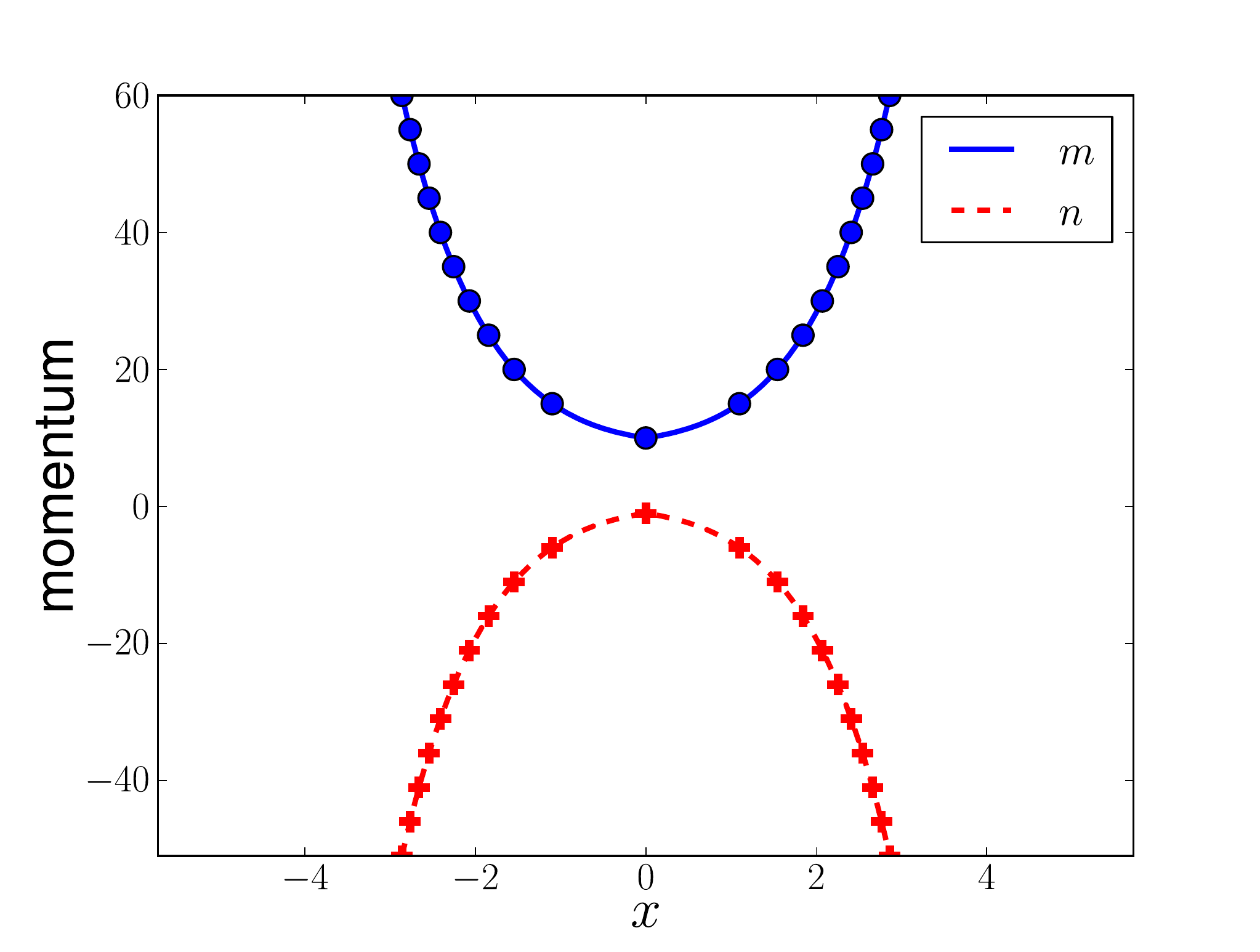} 
\par\end{centering}
\caption{
\label{peakon_orbits2}  Evolution of the velocity (Panel a) and momentum (Panel b) solutions of a peakon-antipeakon coupled pair.\smallskip\\ 
Panel (a) shows the velocity fields of anpeakon-antipeakon
couple at time $t=0$ with $m_{0}=10$, $n_{0}=-1$, $Y_{0}=0$ (solid
lines). The dotted path indicates the locus of the peaks, in the frame
travelling at the mean particle velocity, 
$\dot{X}\left(t\right)=\left(\dot{q}+\dot{r}\right)/2$,
as $t\rightarrow\pm\infty$. Also shown are the velocities at $t=1$.\smallskip\\ 
Panel (b) shows the loci of the $(m,q)$ and $\left(n,r\right)$
particles in a stationary frame of reference centred on the point of collision.
As in Figure  \ref{peakon_orbits}, the dots and crosses indicated
the particles location in phase space at regular time intervals. Here
the best point of comparison is from $t=0$, when the particles are
superposed at $q=r=0$.\smallskip\\ 
Animations showing the time evolution
of both these images are available as supplementary material with
the online copy of this paper.}
\end{figure}
The final step is to consider the mean flow of the system.   This is
given by

\[
\dot{X}=
\left\{
\begin{array}{cc}
\frac{2\mathcal{M}E}{\mathcal{M}^{2}-\left[2Et-\overline{\sgn}\left(Y_{0}\right)\left(n_{0}-m_{0}\right)\right]^{2}}, & t<T_{0}\\
\frac{2\mathcal{M}E}{\mathcal{M}^{2}-\left[2E\left(t-T_{i}\right)-\sqrt{\mathcal{M}^{2}-8E}\right]^{2}} & T_{i}<t<T_{i+1}\end{array}
\right.
\]
  i.e. 
\[
\hspace{-2cm}
X-X_{0}
=
\left\{
\begin{array}{cc}
\left[\tanh^{-1}\left(\frac{2Et-\overline{\sgn}\left(Y_{0}\right)\left(n_{0}+m_{0}\right)}{\mathcal{M}}\right)+\overline{\sgn}\left(Y_{0}\right)\tanh^{-1}\left(\frac{n_{0}-m_{0}}{\mathcal{M}}\right)\right] & t<T_{0}\\
X\left(T_{i}\right)-X_{0}+\left[\tanh^{-1}\left(\frac{2E\left(t-T_{i}\right)-\sqrt{\mathcal{M}^{2}-8E}}{\mathcal{M}}\right)\right.\\
\qquad\qquad\qquad\qquad\qquad\qquad\left.+\tanh^{-1}\left(\frac{\sqrt{\mathcal{M}^{2}-8E}}{M}\right)\right] & T_{i}<t<T_{i+1}\end{array}.
\right.
\]
Figs.  \ref{peakon_orbits}-\ref{peakon_orbits2} illustrate this
analytic solution in the both the peakon-peakon ($E>0$) and peakon-antipeakon
($E<0$) cases. Of particular interest in the peakon-peakon case is
the nature of the collisions that 
occur. These collisions propagate the waltzing peakon couple onwards as
a coherent phenomenon, possessing (in the tunnelling representation)
continuous values for the particle positions, velocities and momenta,
but with a discontinuous first derivative in the rate of change of
momentum.

Note that over one entire cycle of the waltz, the mean speed of propagation
is given by
\[
\overline{\dot{X}}=\frac{1}{T_{i+1}-T_{i}}\int_{T_{i}}^{T_{i+1}}\dot{X}dt=\frac{m_{0}n_{0}}{2\left|m_{0}-n_{0}\right|}\tanh^{-1}\left(\frac{\left|m_{0}-n_{0}\right|}{m_{0}+n_{0}}\right).\]
Using l'Hospital's rule we find that in the limit of zero period oscillations
\[
\lim_{T_{0}\rightarrow0}\overline{\dot{X}}=\lim_{m_{0}\rightarrow n_{0}}\frac{m_{0}^{2}}{2}\left(\frac{m_{0}+n_{0}-\left(m_{0}-n_{0}\right)}{\left(m_{0}+n_{0}\right)^{2}}\right)\left(\frac{1}{1-\left(\frac{m_{0}-n_{0}}{m_{0}+n_{0}}\right)^{2}}\right)=\frac{n_{0}}{2}.\]
This is exactly the velocity of a single peakon solution to the CH
equations, to which the cross-coupled equations collapse under the same
limit.

\paragraph{Triangular compacton couples.}
Similarly, we may consider a compacton couple with a triangular kernel and corresponding Hamiltonian 
\begin{eqnarray}
H=\frac{1}{2}m n (1-a|q-r|)\theta(1-a|q-r|) 
\end{eqnarray}
where $a$ is a constant. For initial conditions $q(0)=r(0)=0$, $n(0)=n_0$, $m(0)=m_0$, $n_0>m_0$
the solution is:
\begin{eqnarray*}
q(t) = \frac{m_0n_0}{2c} t + \frac{m_0^2}{ac^2}(1-e^{-cat/2}) 
,&\quad&
r(t) = \frac{m_0n_0}{2c} t + \frac{n_0^2}{ac^2}(e^{cat/2}-1) \\
m(t) = \frac{cm_0}{m_0+n_0e^{-cat/2} }
,&\quad&
n(t) = \frac{c n_0}{n_0+m_0e^{cat/2} }
\label{eq72}
\end{eqnarray*}%
Here $0\le t\le T_1=({4}/{ac})\ln {n_0}/{m_0}$ where $c=m_0+n_0$ is the conserved momentum. For the second half of the cycle the solution is

\begin{eqnarray*}
\hspace{-1cm}
q(t)  = 
K +\frac{m_0n_0}{2c} t + \frac{n_0^2}{ac^2}(e^{ca(t-T_1)/2}-1) 
\,,&\quad&
r(t) = K +\frac{m_0n_0}{2c} t + \frac{m_0^2}{ac^2}(1-e^{-ca(t-T_1)/2}) 
\,,
\\
m(t) = \frac{c n_0}{n_0+m_0e^{ca(t-T_1)/2} }
\,,&\quad&
n(t) = \frac{c m_0}{m_0+n_0e^{-ca(t-T_1)/2} }
\,,\label{eq72a}
\end{eqnarray*}%
for $T_1<t<2T_1$ and the constant $K=2\frac{m_0 n_0}{ac^2}\ln \frac{n_0}{m_0}+\frac{n_0-m_0}{ac}$ represents the spatial shift during the first half cycle. The full cycle has a period $2T_1=\frac{8}{ac}\ln \frac{n_0}{m_0}$.

\section{The Periodic Equilibrium Solutions} \label{periodicPeakons-sec}

While the 1+1 soliton has only one equilibrium solution, with $m=n$
and $l=0$, the periodic solution has two. The first being the CH
equivalent state with particles super-imposed, and the second with
$m$ and $n$ multipeakons which are a half-step out of phase with each
other. In such cases, the Greens function solution is given in the
interval $-L\leq x<L$ by
\[\hspace{-1cm}
m(x,t)=\sum_{a=1}^{M}m_{a}\left(t\right)\delta\left(x-q_{a}\left(t\right)\right),\qquad u(x,t)=\sum_{a=1}^{M}m_{a}\left(t\right)\frac{\cosh\left(L-|x-q_{a}\left(t\right)|\right)}{4\sinh L},\]
 \[\hspace{-1cm}
n(x,t)=\sum_{b=1}^{N}n_{b}\left(t\right)\delta\left(x-r_{b}\left(t\right)\right),\qquad v(x,t)=\sum_{b=1}^{N}n_{b}\left(t\right)\frac{\cosh\left(L-|x-r_{b}\left(t\right)|\right)}{4\sinh L},\]
and by the periodicity conditions, 
$m(x+2L)=m(x),\  n(x+2L)=n(x),$ etc., otherwise. 
 
 For the case $M=N=1$, upon dropping the indices
$a$ and $b$ and defining once again
\[
Y(x)=q(x)-r(x),\]
 we see (eventually) that
 \[
\frac{d^{2}}{dt^{2}}2\tan^{-1}\left(\tanh\frac{L-|Y|}{2}\right)
=
-2E\,\mbox{sgn}^{2}\left(Y\right)\tanh\left(L-|Y|\right).
\]
From this relation, we obtain the quadrature
 \[
\frac{d|Y|}{dt}=\left(\frac{1}{2}\ln\left(\cosh\left(L-|Y|\right)\right)+C\right)e^{-\frac{E\cosh\left(L-|Y|\right)}{\sinh\left(L\right)}}
,
\]
 where as before
 \[
E=\frac{m_{0}n_{0}}{2}e^{-|Y_{0}|},\]
and the constant of integration $C$ is found from the initial condition
 \[
\left.\frac{d|Y|}{dt}\right|_{t=0}
=
\sgn\left(Y_{0}\right)\left(n_{0}-m_{0}\right)\frac{\cosh\left(L-|Y_{0}|\right)}{4\sinh L}
\,.
\]

\section{The 2+2 Peakon-Antipeakon solution} \label{ppbar-solution-sec} 

\begin{figure}[h!]
\begin{centering}
\includegraphics[clip,width=0.85\textwidth]{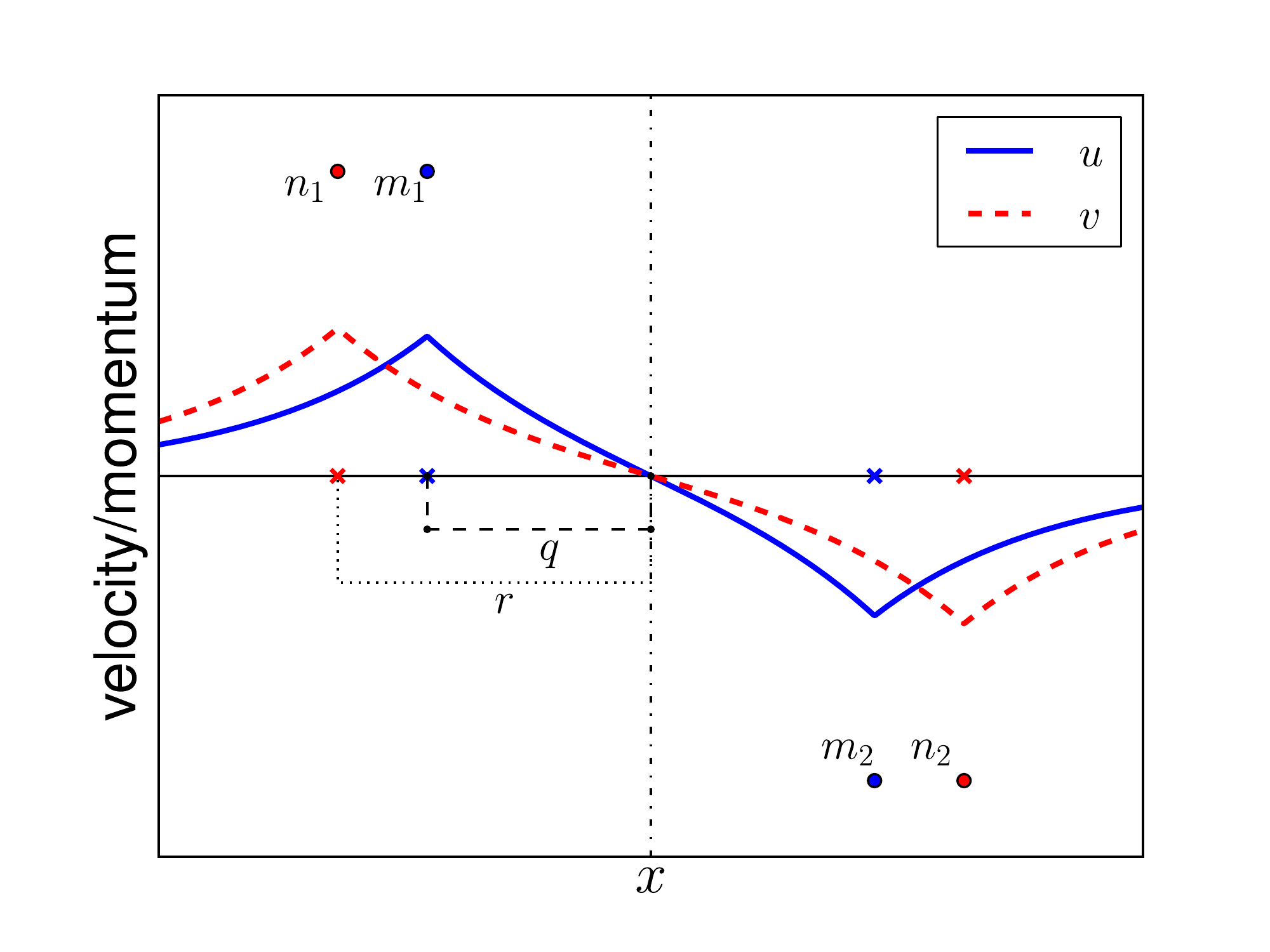} 
\par\end{centering}

\caption{\label{Headon_Collision}This is the head-on collision set-up used in our
analysis. It has a rightward travelling peakon-peakon couple in the left
half plane and an antisymmetric counterpart on the right. Each particle
couple undergoes {}``waltzing'' collisions before reaching, by symmetry,
a singular collision at the point $x=0$.\smallskip\\ 
An animation that shows the
time evolution of such a peakon-antipeakon collision is available
as supplementary material with the online copy of this paper.}

\end{figure}

We consider now the anti-symmetric collision of two couples of peakons in
which, using the labelling pattern indicated in Figure \ref{Headon_Collision}
we have the following reduction of the system (\ref{qa})--(\ref{nb}):
\begin{eqnarray*}
m_{1}  =-m_{2}\equiv m,\quad
n_{1}  =-n_{2}\equiv n, \quad
q_{1}  =-q_{2}\equiv q, \quad
r_{1}  =-r_{2}\equiv r.
\end{eqnarray*}
The reduced dynamical system in this case has the canonical Hamiltonian form,
\begin{eqnarray*}
\dot{q} = \frac{\partial H}{\partial m} 
& = & \frac{1}{2}n\left(e^{-|q-r|}-e^{-|q+r|}\right)
,\label{qeq}
\\
\dot{r} =
 \frac{\partial H}{\partial n}
& = & \frac{1}{2}m\left(e^{-|q-r|}-e^{-|q+r|}\right)
,\label{req}
\\
\dot{m} =  -\,\frac{\partial H}{\partial q}
& = &  \frac{1}{2}mn\left(\sgn(q-r)e^{-|q-r|}-\sgn(q+r)e^{-|q+r|}\right)
,\label{meq}
\\
\dot{n} =
-\,\frac{\partial H}{\partial r}
& = & -\frac{1}{2}mn\left(\sgn(q-r)e^{-|q-r|}+\sgn(q+r)e^{-|q+r|}\right)
.\label{neq}
\end{eqnarray*}
The Hamiltonian for this system is 
\[
H=\frac{1}{2}mn\left(e^{-|q-r|}-e^{-|q+r|}\right)\]
 and its canonical Poisson bracket arises from the symplectic form 
 $\omega = dq\wedge dm +  dr\wedge dn$.

We choose to consider the colliding case
\[
\sgn\left(n\right)=\sgn\left(m\right)>0
\]
and 
\[
\sgn\left(q\right)=\sgn\left(r\right)<0.
\]

Initially we suppose that $q(0)=r(0)=\ln k<0$ where from the energy
conservation \[
k=\sqrt{1-\frac{2H}{m_{0}n_{0}}},\]
 $m_{0}=m(0)$ and $n_{0}=n(0)$. Then we assume that $0>q(t)>r(t)$
until after some time $T$ again $q(T)=r(T)$ and another 'exchange'
of the positions of $n$ and $m$ takes place. Under these circumstances
the system (\ref{qeq})--(\ref{neq}) for $0\le t\le T_{1}$ has the form
\begin{eqnarray*}
\dot{q} = -ne^{r}\sinh q
\label{qeq0}
\,,\quad
\dot{r} = -me^{r}\sinh q
\label{req0}
\,,\quad
\dot{m} = mne^{r}\cosh q
\label{meq0}
\,,\quad
\dot{n} = mne^{r}\sinh q
\label{neq0}
\end{eqnarray*}
 The system conserves the energy $H=-mne^{r}\sinh q$ and can be integrated,
giving:
\begin{eqnarray}
q(t) = \ln\left(\frac{(k+1)+(k-1)e^{-n_{0}kt}}{(k+1)-(k-1)e^{-n_{0}kt}}\right),\label{qeq1}
&\quad&
r(t) = \ln\left(\frac{kn_{0}}{n_{0}-Ht}\right),\label{req1}\\
m(t) = H\frac{(k+1)^{2}-(k-1)^{2}e^{-2n_{0}kt}}{2kn_{0}(1-k^{2})e^{-n_{0}kt}},\label{meq1}
&\quad&
n(t) = n_{0}-Ht.\label{neq1}
\end{eqnarray}
 The time to the next exchange, $T_{1}$ can be determined from $q(T_{1})=r(T_{1})$,
i.e. from the equation

\begin{equation}
n_{0}-HT_{1}=kn_{0}\frac{(k+1)-(k-1)e^{-n_{0}kT_{1}}}{(k+1)+(k-1)e^{-n_{0}kT_{1}}}.\label{T1eqn}\end{equation}

\noindent If $n_{0}>m_{0}$ this equation has a real root, which is
less or equal to $n_{0}/H$. For $0<t<T_{1}$ $n(t)$ is decreasing
and $m(t)$ is increasing.

The equations for the time following the exchange, i.e. when $0>r(t)>q(t)$
are
\begin{eqnarray*}
\dot{q} = -ne^{q}\sinh r,\label{qeq2}
&\quad&
\dot{r}  =  -me^{q}\sinh r,\label{req2}\\
\dot{m} = mne^{q}\sinh r,\label{meq2}
&\quad&
\dot{n}  =  mne^{q}\cosh q.\label{neq2}
\end{eqnarray*}
with solutions
\begin{eqnarray*}
q(t) =  \ln\left(\frac{k_{1}m_{1}}{m_{1}-H(t-T_1)}\right),\label{qeq3}
&\quad&
r(t)  =    \ln\left(\frac{(k_{1}+1)+(k_{1}-1)e^{-m_{1}k_{1}(t-T_1)}}{(k_{1}+1)
-(k_{1}-1)e^{-m_{1}k_{1}(t-T_1)}}\right),\label{req3}\\
m(t) = m_{1}-H(t-T_1),\label{meq3}
&\quad&
n(t)  =  H
\frac{(k_{1}+1)^{2}-(k_{1}-1)^{2}e^{-2m_{1}k_{1}(t-T_1)}}{2km_{1}(1-k_{1}^{2})
e^{-m_{1}k_{1}(t-T_1)}},\label{neq3}
\end{eqnarray*}
 where $m_{1}=m(T_{1})$, $n_{1}=n(T_{1})$ from (\ref{meq1})  and (\ref{neq1})
 are the new (updated) initial conditions, and the updated
value for the former constant $k$ is now 
\[
k_{1}=\sqrt{1-\frac{2H}{m_{1}n_{1}}}\,.\]

These solutions are valid until the time of the next collision, $T_1\le t\le T_{2}$
when again $q(T_{2})=r(T_{2})$ or, $T_{2}$ is a solution a similar
transcendental equation:

\begin{equation}
m_{1}-H(T_{2}-T_1)=k_{1}m_{1}\frac{(k_{1}+1)-(k_{1}-1)
e^{-m_{1}k_{1}(T_{2}-T_1)}}{(k_{1}+1)+(k_{1}-1)e^{-m_{1}k_{1}(T_{2}-T_1)}}
,
\label{T2eqn}
\end{equation}

The total time of the two exchanges, i.e. for a full cycle is $T=T_{2}$.

\begin{figure}[t]

\begin{centering}
(a) \includegraphics[clip,width=0.45\textwidth]{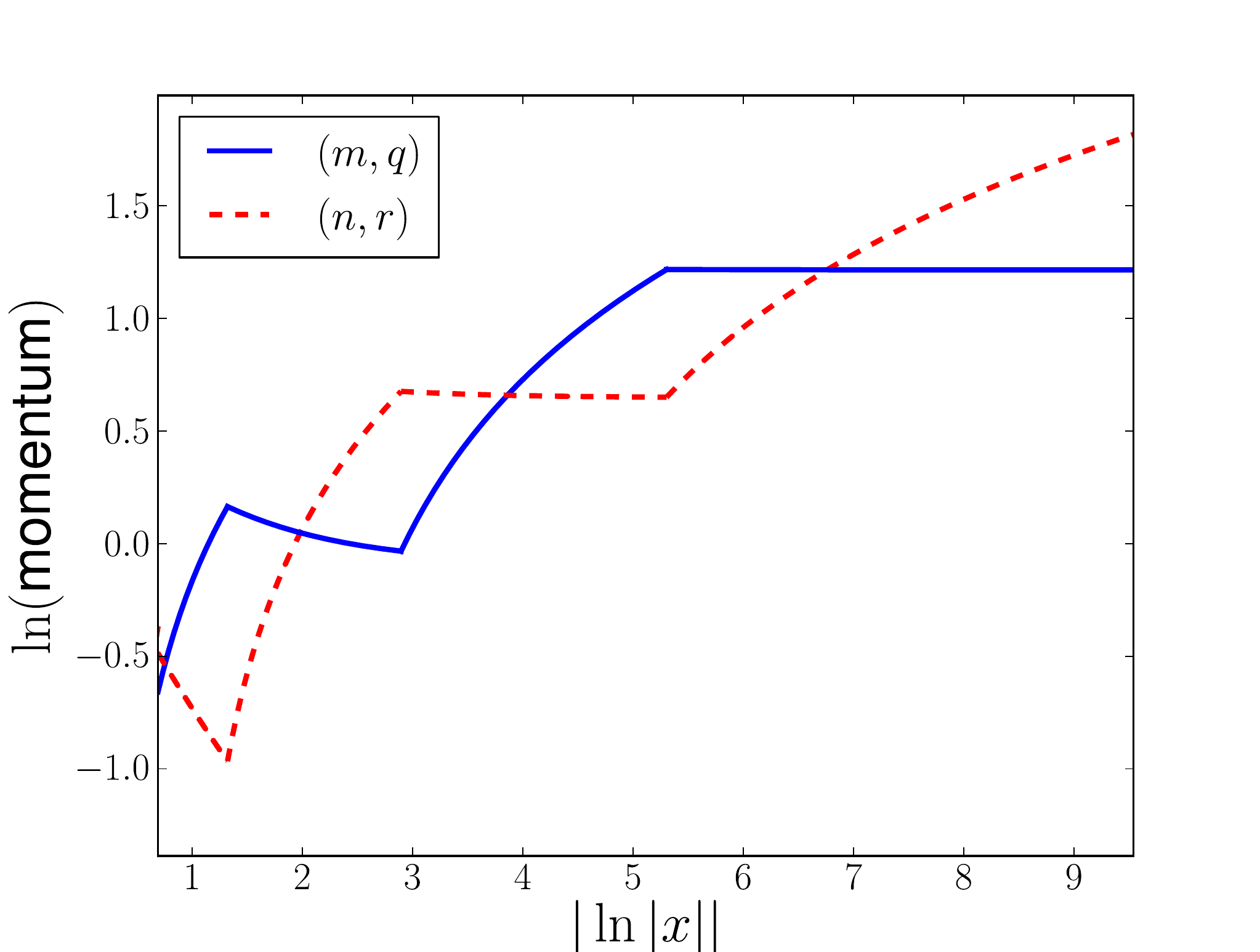}
(b)\includegraphics[clip,width=0.45\textwidth]{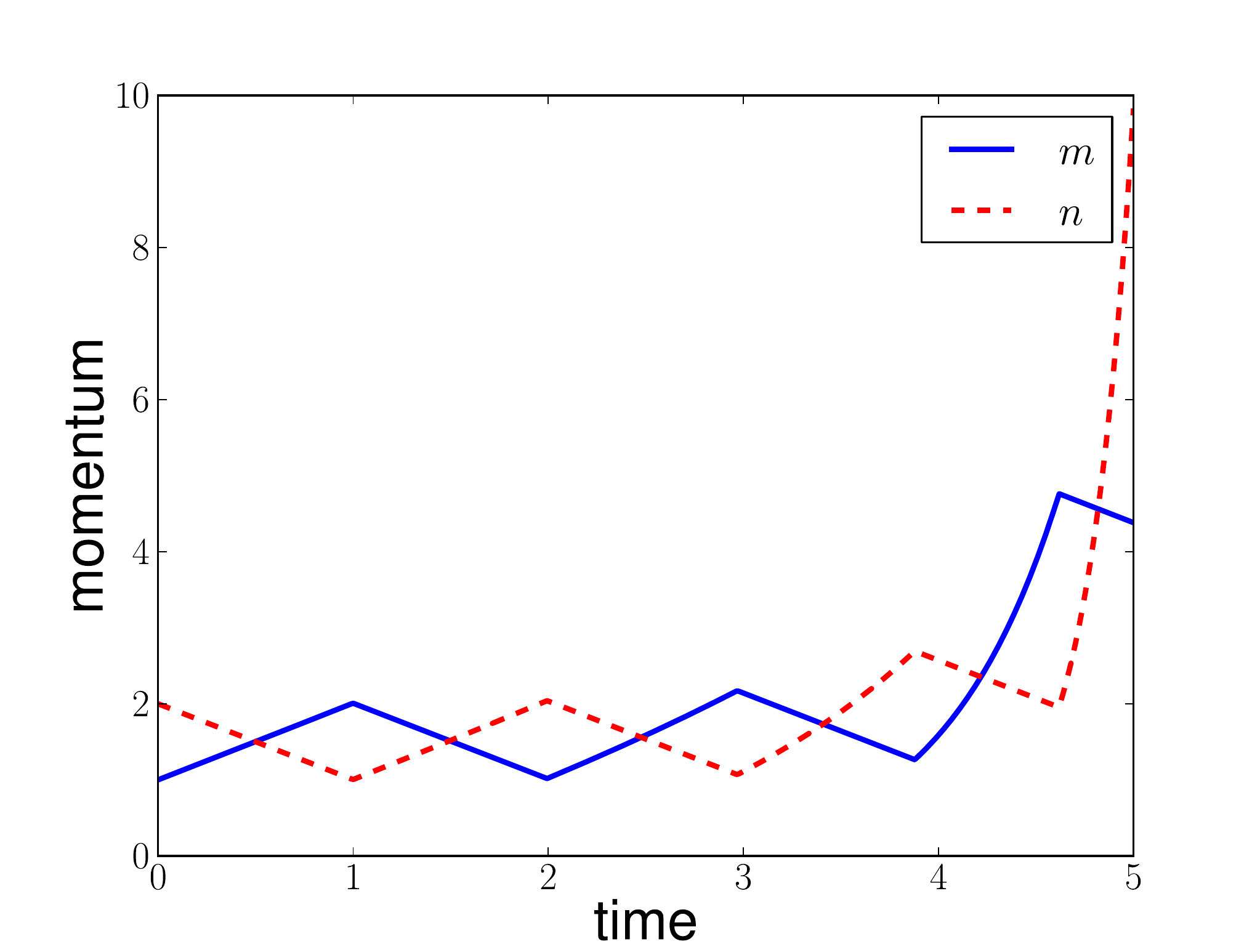} 
\end{centering}

\caption{\label{peakon_peakon}  Evolution of the momentum of the peakon couple in the anti-symmetric headon collision in terms of position (Panel a) and time (Panel b).\smallskip\\ 
Panel (a) shows a log-log plot of
momentum versus position for the peakon couple in the peakon-antipeakon
collision, calculated from a numerical integration of the governing
ODEs (\ref{qeq0})-(\ref{neq0}), (\ref{qeq2})-(\ref{neq2}). Here
$m\left(0\right)=1,n\left(0\right)=2,q\left(0\right)=r\left(0\right)=-2$.
This plot illustrates the evolution over two complete cycles of the
{}`waltzing' peakon, while also showing the near exponential increase
in the momentum amplitudes over each full cycle.\smallskip\\ 
Panel (b) shows
momentum versus time for similar data, with $m\left(0\right)=1,n\left(0\right)=2,q\left(0\right)=r\left(0\right)=-3$.
The last stage shows the rapid change from behaviour very similar to the
pure peakon couple to near exponential growth as the particles in the peakon couple approach
each other. In both cases the behaviour of the antipeakon couple is identical modulo a reversal of signature.}

\end{figure}

Thus the evolutionary process of the system that emerges from these 
solutions between the collisions is an iterative map illustrated
in Figure \ref{peakon_peakon}.

\section{Other Collisions} \label{othercollisions-sec}

We now present the results of numerical integration of the interaction
between a peakon couple and other coherent structures. The numerical
code solves the ODEs for the particle momenta and peak locations using
an implicit midpoint rule integrator and a collision detection routine
to identify the points of discontinuity in the momentum forcing term.
This method can be used to investigate the short term behaviour of
the ($M,N$) peakon equations for cases with $M$ and/or $N$ greater
than one, for which we do not yet have an analytic description.

\begin{figure}[t]
\begin{centering}
(a) \includegraphics[clip,width=0.42\textwidth]{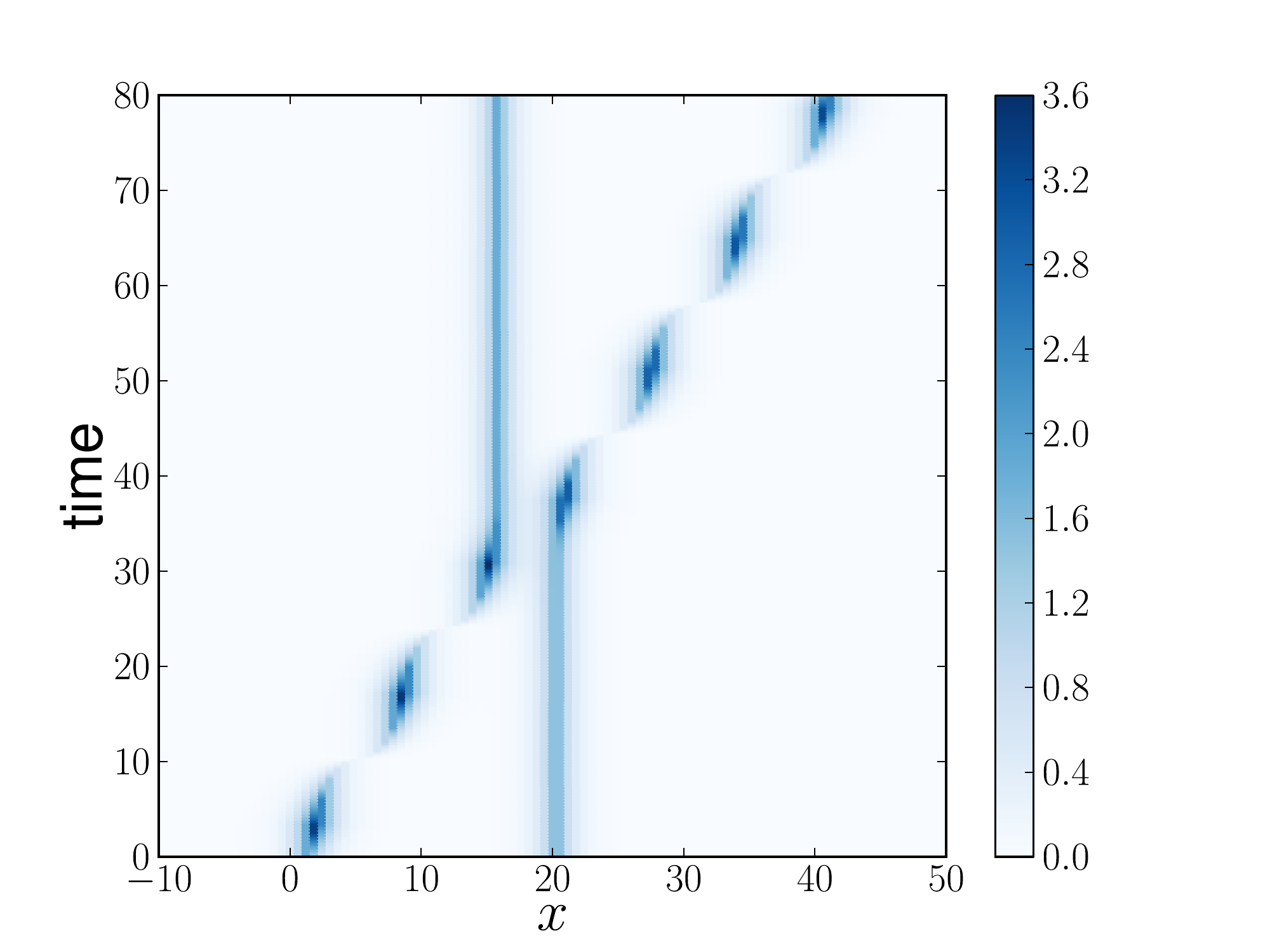} 
(b) \includegraphics[clip,width=0.42\textwidth]{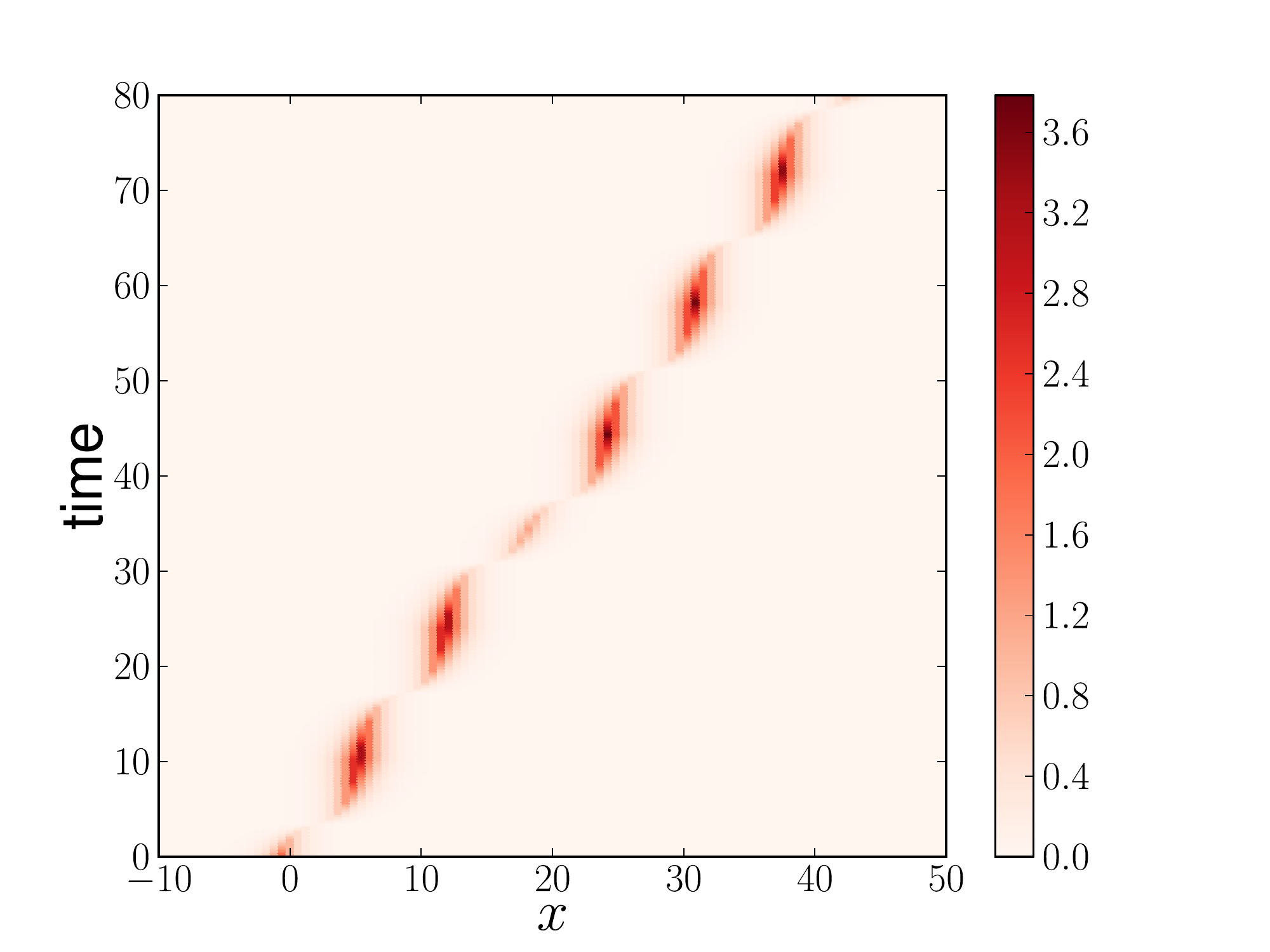} 
\end{centering}

\caption{\label{u1-v1} Spacetime plots showing an example of the evolution
of the $u$ (panel a) and $v$ (panel b)  velocity fields, under the interaction of a peakon
couple with a free resting soliton. Here all particles are initially
of momentum $m_{1}=m_{2}=n_{1}=4$, with initial positions $q_1=1$, $q_2=20$,
$r_{1}=-1$. Comparison of the panels shows the beating amplitude and phase
locking of the waltzing couple, which is interrupted when the $n$ particle
interacts with the second $m$ particle.}

\end{figure}

\begin{figure}[t]
\begin{centering}
\includegraphics[clip,width=0.45\textwidth]{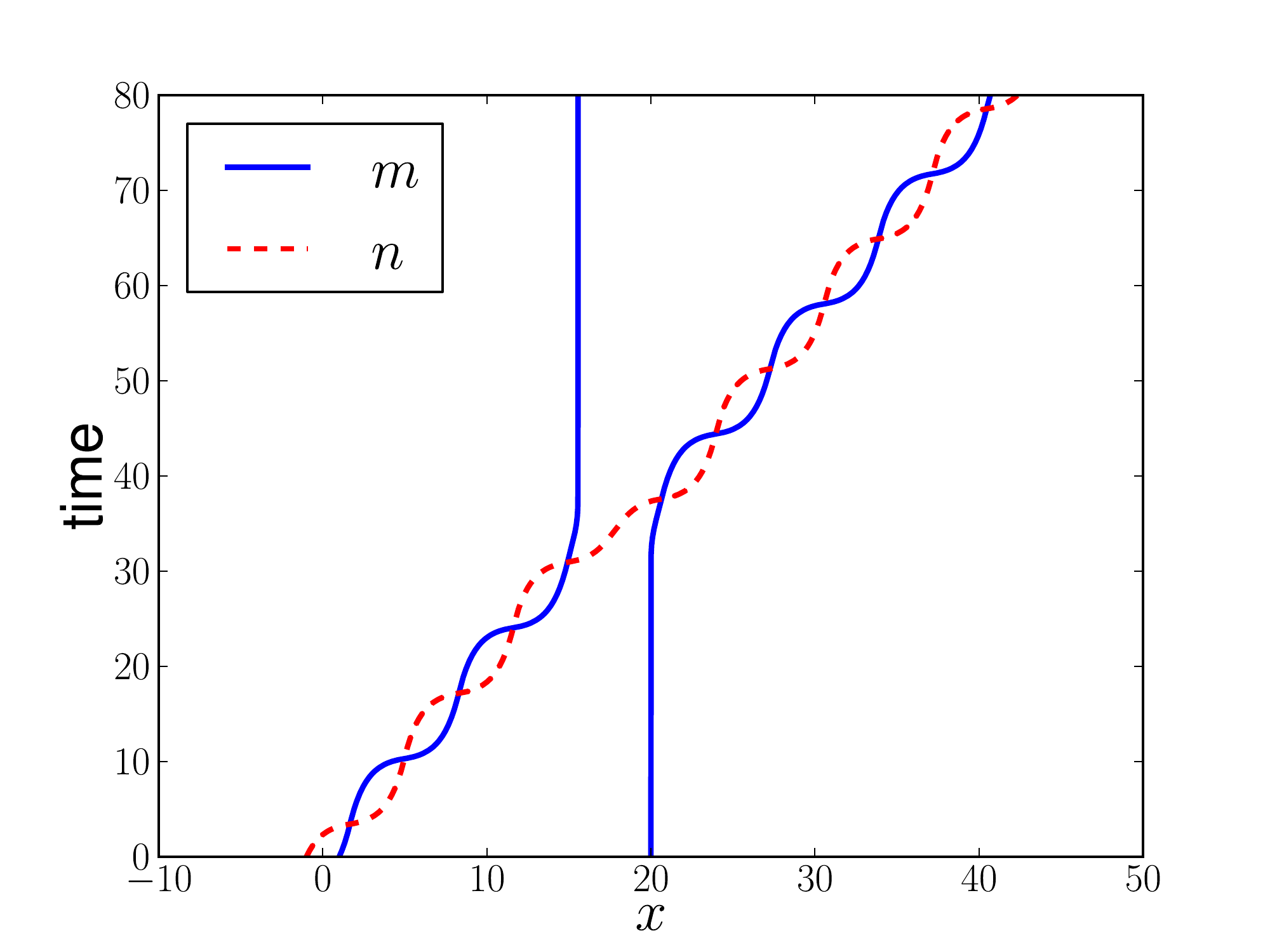} 

\par\end{centering}

\caption{\label{mn} The space-time track of the particles presented in Figure
\ref{u1-v1}. One waltzing partner deserts its old partner on interacting
with another particle of opposite type, with a barely perceptible
change in the period of oscillation (tempo) in the waltzing couple.\\
Animations that show the time evolution of this image and the underlying peakons
are available as supplementary material with the online copy of this
paper.}

\end{figure}

\begin{figure}[t]
\begin{centering}
(a) \includegraphics[clip,width=0.42\textwidth]{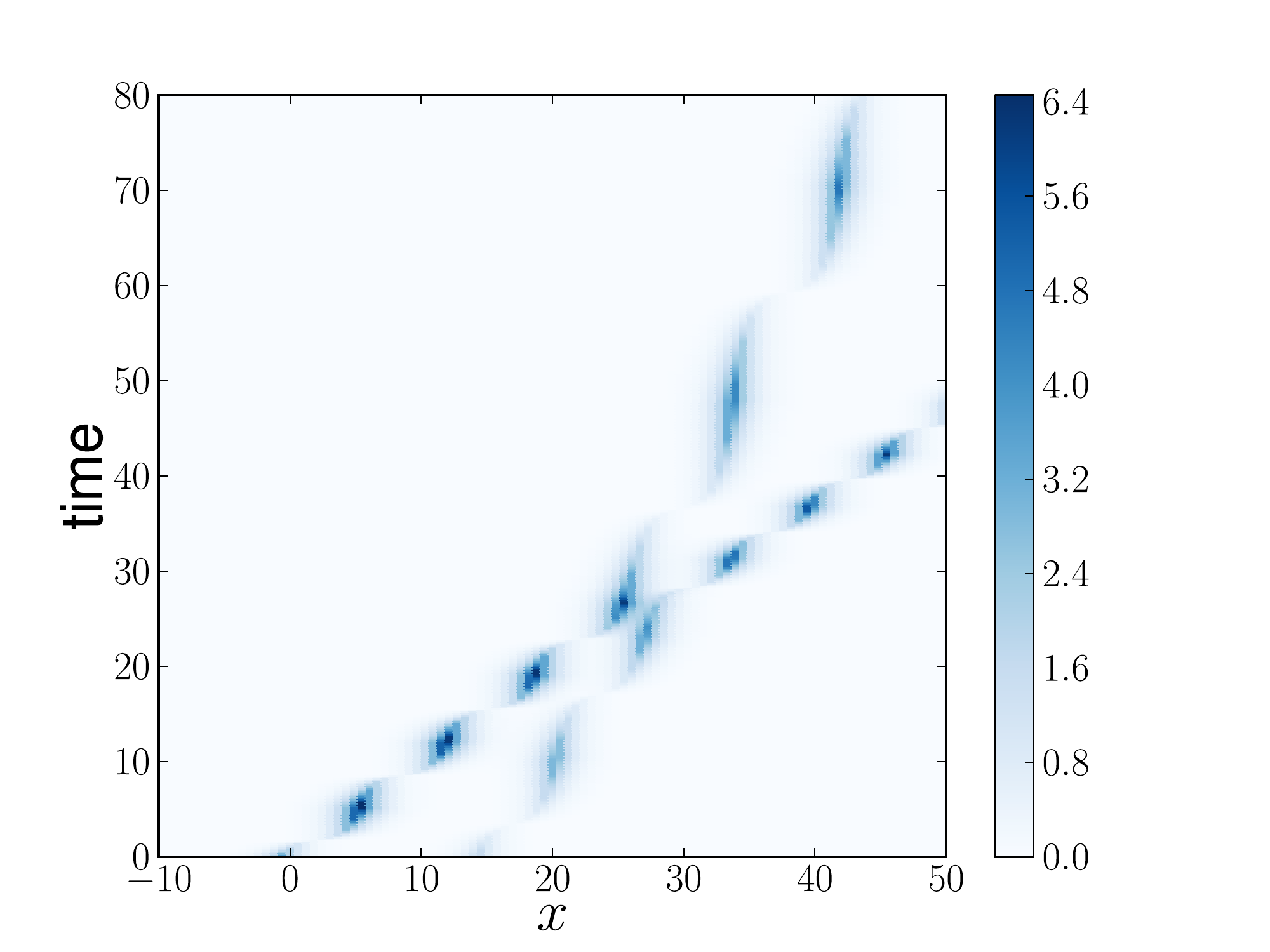}
(b) \includegraphics[clip,width=0.42\textwidth]{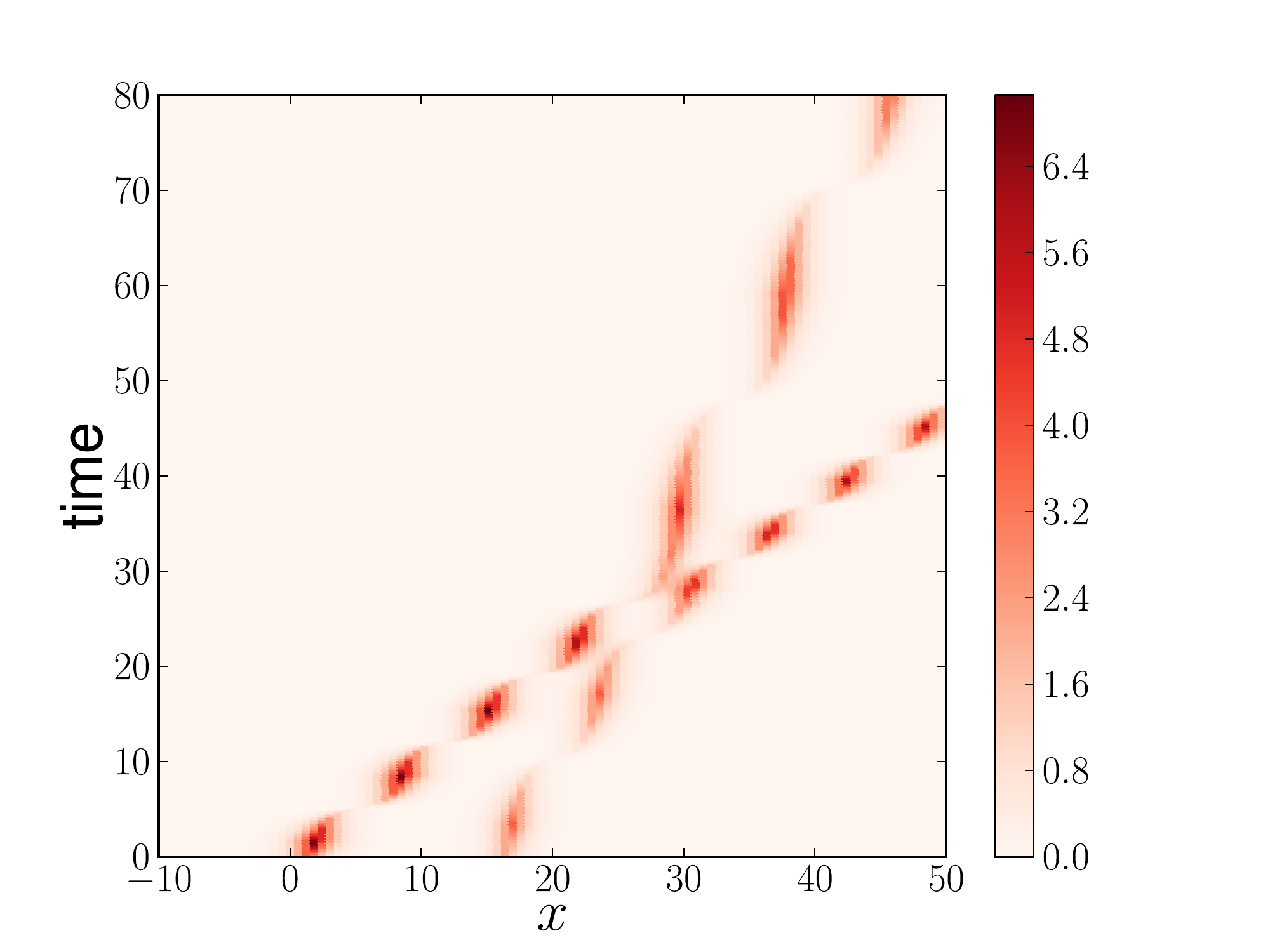} 
\end{centering}

\caption{\label{u2-v2} Spacetime plots showing an example of the evolution
of the $u$ (Panel a) and $v$ (Panel b) velocity fields, under the interaction of two
waltzing couples of peakons. Here $m_{1}=n_{1}=8$, $m_{2}=n_{2}=4$, $q_{1}=-1$,
$q_{2}=14$, $r_{1}=1$, $r_{2}=16$. The peakon couples interact and
exchange some momentum leading two a speeding up of the rightmost
couple. However the collision also affects the {}``internal oscillation''
energy of the two couples of particles, changing the period of oscillation
of both constructs. }

\end{figure}

\begin{figure}[t]
\begin{centering}
\includegraphics[clip,width=0.45\textwidth]{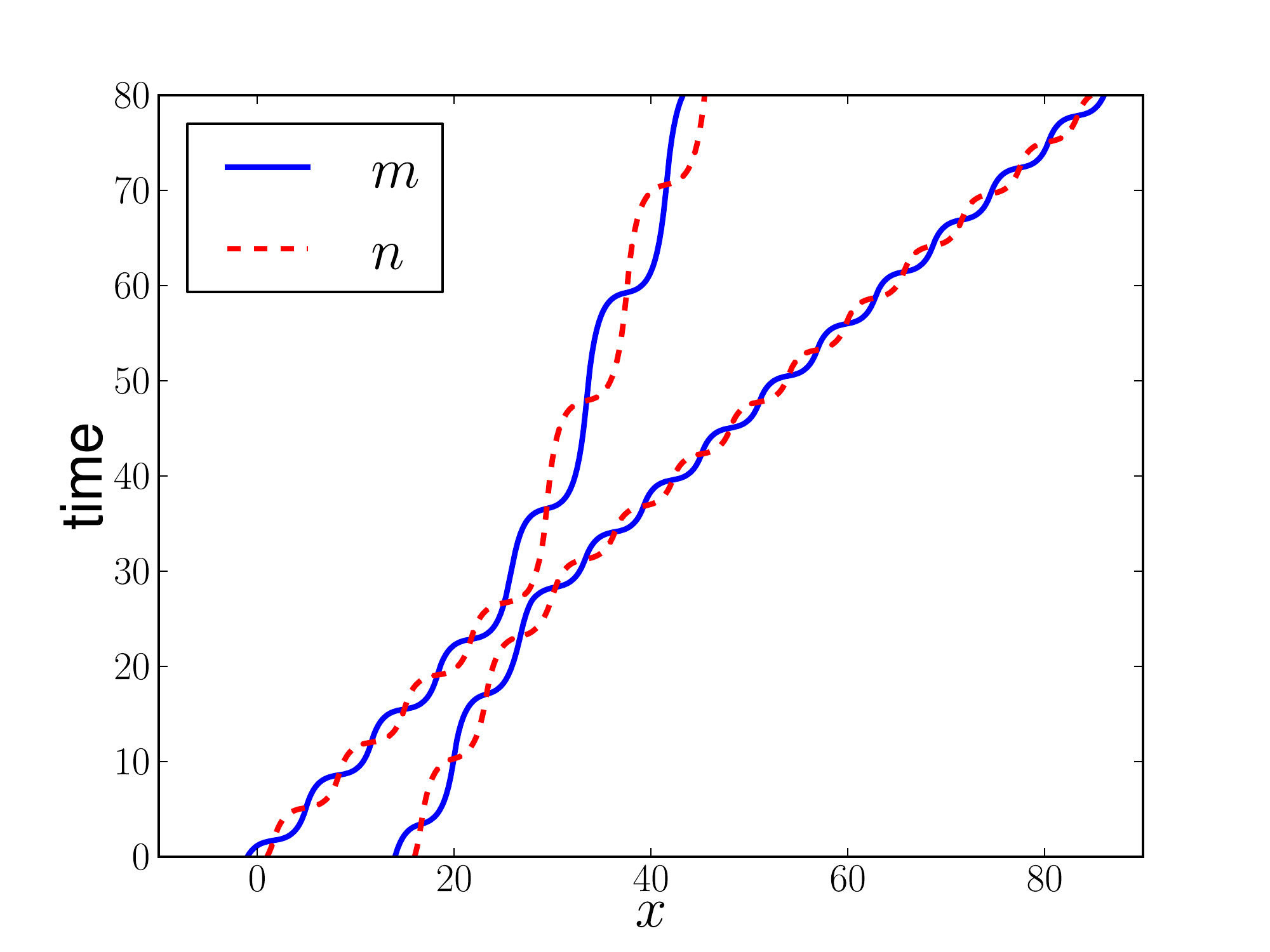} 

\par\end{centering}

\caption{\label{mn2} The space-time tracks of the waltzing couples of peakons from Figure \ref{u2-v2}. This
makes it clearer that the two waltzing couples exchange momentum without ever
actually overlapping, as well as illustrating the increased period
of the lefthand couple and decreased period of the righthand one. Although
not obvious, the post-exchange mean propagation speeds of the two
couples also differ from those pre-collision.\\
Animations showing the
time evolution of this image and the underlying peakons are available
as supplementary material with the online copy of this paper.}

\end{figure}

\subsection{Interaction between a peakon couple and a solitary peakon}

This is the case with $M=2$, $N=1$, $\sgn\left(m_{1}\right)=\sgn\left(m_{2}\right)=\sgn\left(n_{1}\right)$ in
which a peakon couple interacts with a single peakon. 
The results of a typical integration are shown
in Figs.  \ref{u1-v1}-\ref{mn}, in terms of both the generated
velocity fields and the tracks of the particles. Since the effect
of one peakon on another falls exponentially with distance, a tightly
bound peakon-peakon couple feels little influence from a distant solitary
peakon, and by the same measure causes little change in the solitary
peakon itself. However once the couple cycles sufficiently near at hand
for the distance of the solitary peakon from the couple to be comparable
to the maximum separation distance of the couple, the particle feels
sufficient effect to be swept into motion. The ordering condition
on the line is sufficient to guarantee that the formerly solitary
particle is never overtaken by the other member of its type. In
all integrations performed, the solitary peakon is swept up into a
new waltzing couple, with the former partner abandoned, however we have
no definite proof that there does not exist a parameter regime in
which the three particles may form a coherent propagating structure.

\subsection{Overtaking between two peakon couples}

This is the case with $M=2$, $N=2$, $\sgn\left(m_{a}\right)=\sgn\left(n_{b}\right)$,
in which two peakon couples interact. In general this requires the mean
propagation speed of the trailing couple to be larger than that of the
leading couple, so that the two structures interact in a manner equivalent
to the CH overtaking collision. The results of a typical integration
are shown in Figs.  \ref{u2-v2}-\ref{mn2}. As in the case of interaction
with a solitary peakon, the particles behave in the manner of two
pure cycling peakon-peakon couples, so long as the separation of the
centres of the couples is significantly longer than the the maximum
separation distance within either couple. Within the parameter space
explored, the subsequent interaction is always at a distance, with
no extra entanglement of the loops of the two tracks. The resulting
exchange of momenta leads to an acceleration of the initially slower
leading couple, with concomitant deceleration of the trailing couple.
However, this exchange is not perfect, with additional energy partitioned
into or out from the oscillatory motion, above that necessary to change
the speed of progression. From the observed behaviour it can be conjectured
that the long time solution of the ($n_{1}+n_{2}$,$n_{1}$) peakon
cross-coupled equations for $n_{1},n_{2}\in\mathbb{N}$ is to form a chain
of $n_{1}$ peakon-peakon couples, well ordered by their mean propagation
speed, with $n_{2}$ near non propagating solitary peakons remaining.
Due to the complex nature of propagation speed formula it is not,
however possible to guarantee well ordering in either the momenta
or the separation distances involved.

\section{CCEP coupled compacton solutions}

\label{compactons-sec}

\begin{figure}
\begin{centering}
\includegraphics[width=0.45\textwidth]{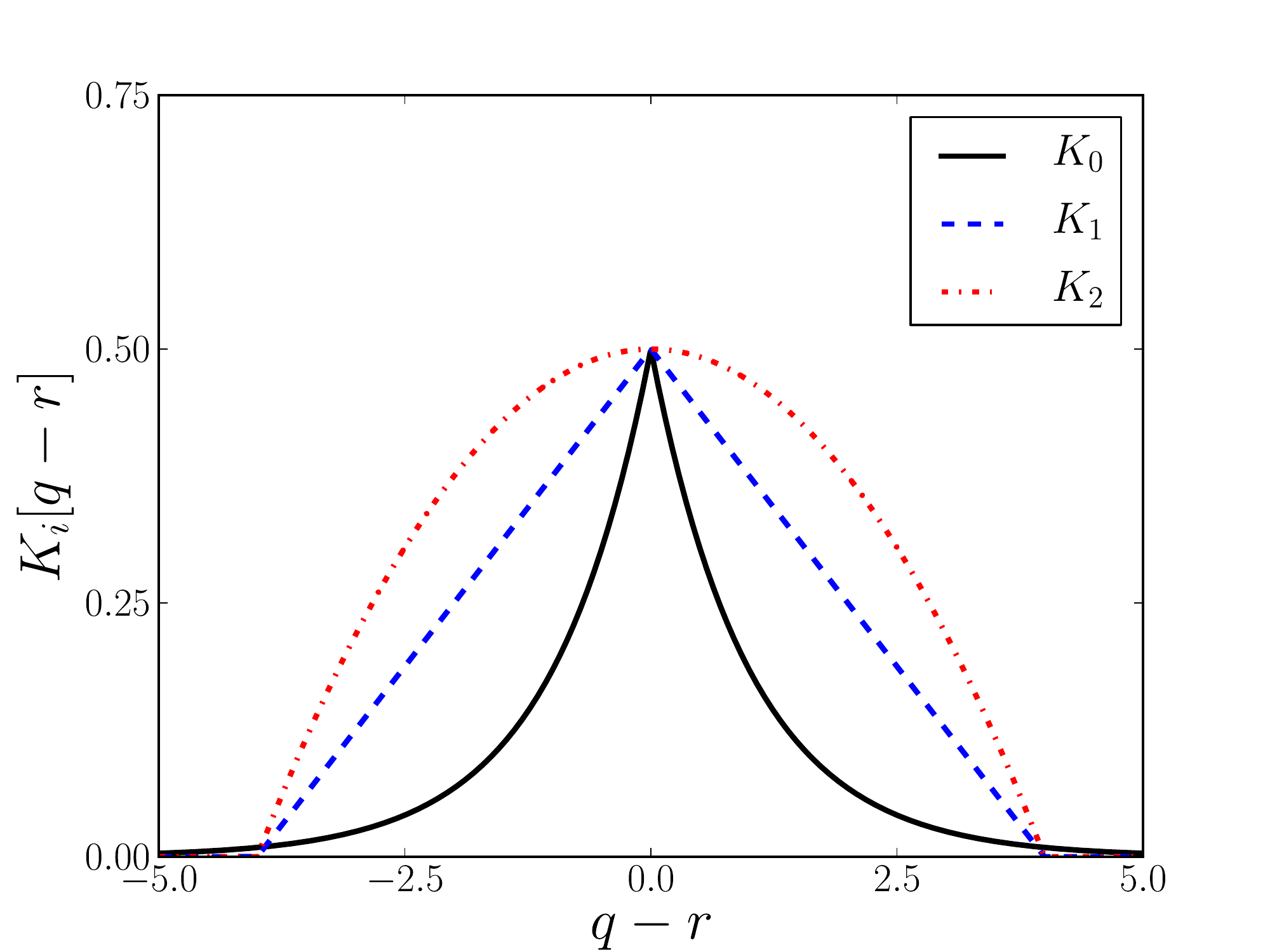}
\par\end{centering}

\caption{\label{fig:kernel_profiles}Plot shows the profile of the triangular and parabolic kernels,
$K_{1}$ and $K_{2}$, defined in equations (\ref{eq:triangle_kernel}) and (\ref{eq:parabolon_kernel}),
as well as the CCCH peakon profile, $K_{0}$. The parabolic profile, $K_{2},$ has discontinuities in derivatives
at $x = \pm 4$ but is smooth across the origin. The triangular profile,
$K_{1}$, has discontinuities in its derivative at all three points,
$x\in[-4,0,4]$.}

\end{figure}

An arbitrary norm on the tangent space of vector fields,
 $\left\Vert u\right\Vert =\left(u,\mathcal{G} u\right),$
with $\mathcal{G}$ a symmetric operator on $u$ and its derivatives,
induces cross-coupled equations through application of the CCCH framework
to the new Lagrangian
$
l_{\mathcal{G}}\left(u,v\right) =\left(u, \mathcal{G}v\right).
$
 In particular, the convolutions 
$
u = K_\mathcal{G}*m
\quad\hbox{and}\quad
v = K_\mathcal{G}*n,
$
define the kernel $K_\mathcal{G}$ for the system, which inverts the momentum-velocity relation in the definitions of the momenta in the Legendre transformation,
\[
m=\frac{\delta l_{\mathcal{G}}}{\delta v}
\quad\hbox{and}\quad
n=\frac{\delta l_{\mathcal{G}}}{\delta u}.
\]
The numerical method, used in the previous section to integrate the Hamiltonian ODEs governing the peakon
solutions of CCCH, was extended to solve
the system of evolution equations for general Hamiltonians of the form
\begin{equation}
\label{eqn-gen-ham}
\frac{1}{2}H=\sum_{a,b=1}^{M,N}m_{a}(t)n_{b}(t)K_\mathcal{G}(|q_{a}-r_{a}|).
\end{equation}
We consider in particular the two
\emph{compactly supported} kernels,
\begin{eqnarray}
K_{1}(x)  =\tfrac{1}{2}\left(1-\frac{x}{4}\right)
\theta\left(1-\frac{x}{4}\right)
\label{eq:triangle_kernel}
\quad\hbox{and}\quad
K_{2}(x)  =\tfrac{1}{2}\left(1-\frac{x^{2}}{16}\right)\theta\left(1-\frac{x}{4}\right)
,
\label{eq:parabolon_kernel}
\end{eqnarray}
that are, respectively, triangular and parabolic. The profiles of these two compacton kernels are displayed in Figure 
\ref{fig:kernel_profiles}, in comparison with the peakon profile,  $\frac12\exp(-|x|)$. 

Much of the analysis of Section \ref{coupledPpair-sec} generalises immediately 
to singular solutions of cross-coupled Euler-Poincar\'e (CCPE) equations with compacton kernels.
 When the centres of two solitary $m$ and $n$ particles lie within each
 other's domains of support,  the particles propagate together with the
 same form of cyclical motion for CCPE compactons as exhibited by the CCCH peakons. For example, numerical integrations shows the same results as for the peakon in the interaction of a cyclically propagating compacton couple with a solitary, stationary compacton, for each of the kernels, $K_{1}$ and $K_{2}$. These cases are essentially the same as in Figure  \ref{mn} for identical initial particle weights and separations. 

\begin{figure}[h!]
(a)\includegraphics[width=0.45\textwidth]{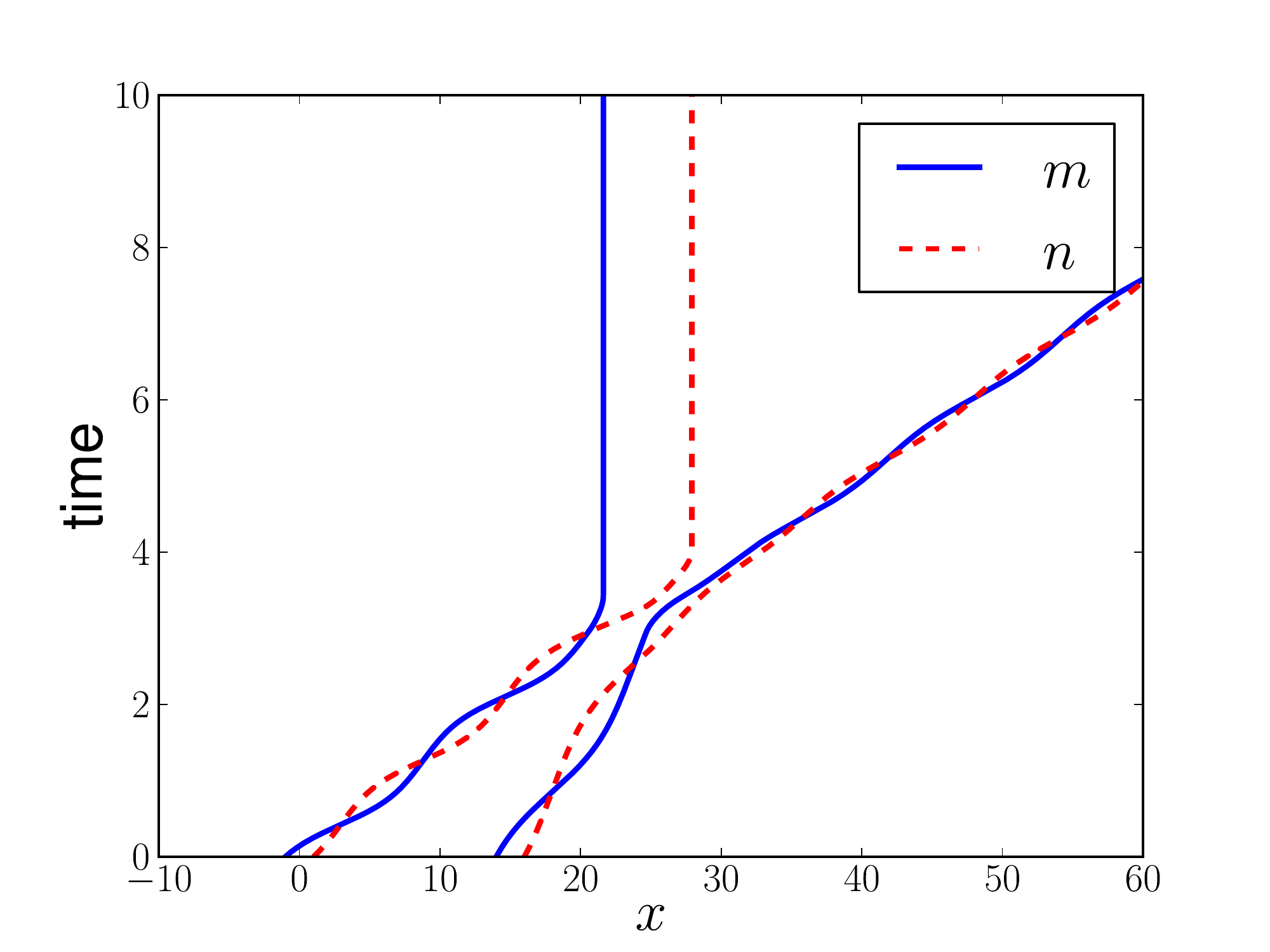}
(b)\includegraphics[width=0.45\textwidth]{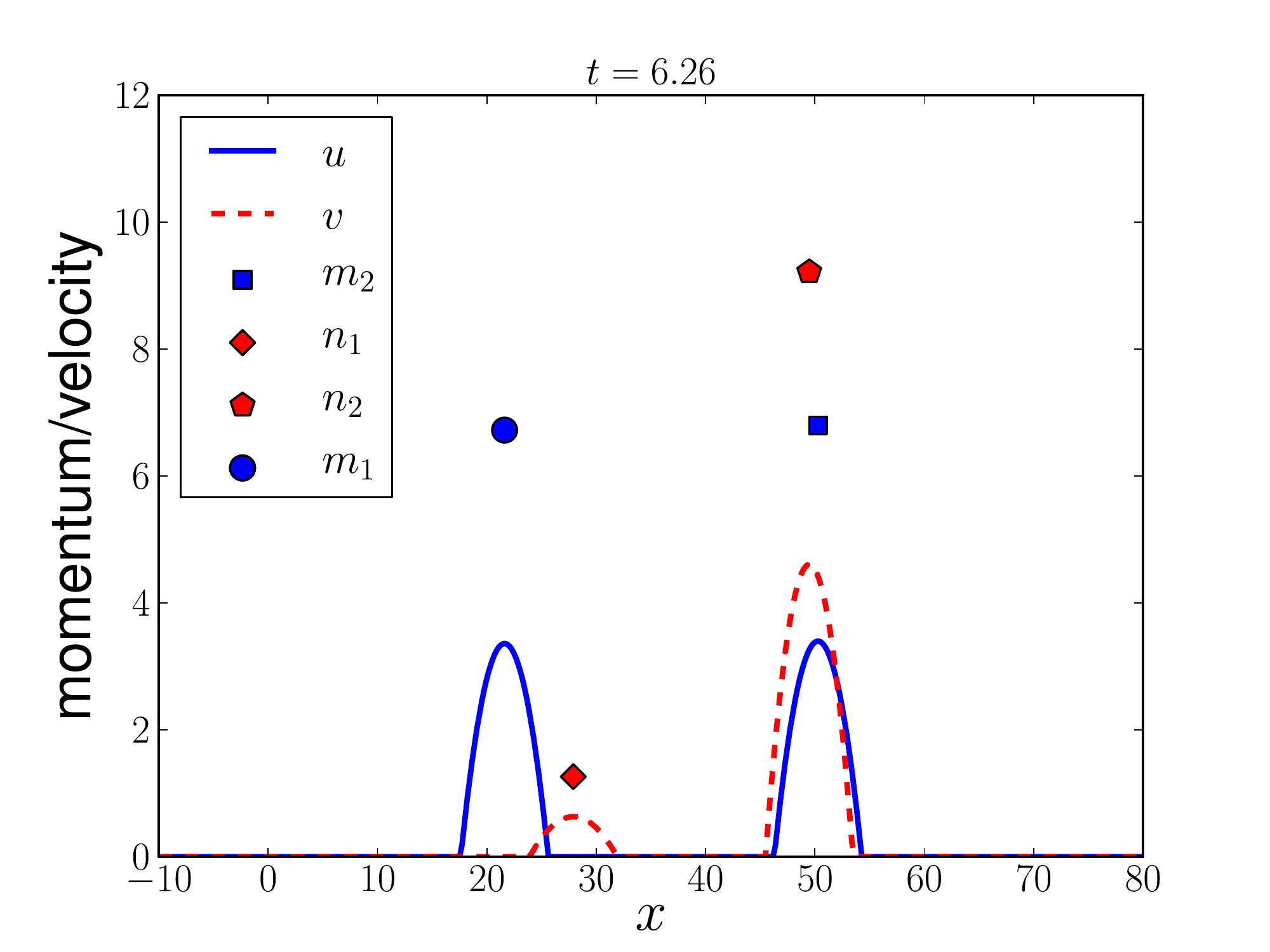}

\caption{\label{fig:mn_compacton_collisions} Panel (a) depicts  spacetime plots for the pairwise
overtaking problem equivalent to Figure \ref{mn2}, but with the parabolic compacton
profiles in equation (\ref{eq:parabolon_kernel}). 
The interaction is sufficiently violent that the trailing $n$ (blue)
particle gains enough extra velocity to escape the region of support
of its partner (red), breaking the cyclic red-blue couple into two stationary particles.\\
Panel (b) shows the positions and weights of the particles shortly after the collision, as well as the resultant velocity profiles. This clearly shows that the separation of the particles exceeds the compacton width.\\
This behaviour was also found for other compacton kernels (such as triangular compactons) in regimes
with sufficiently energetic compacton couples. Overtaking collisions of cross-coupled compactons
 can thus influence the form of the long time
solutions. This behaviour is not seen for systems of compacton equations that contain only
a self interaction term.\\
Animations that show the time evolution of this image are available as supplementary material with the online copy of this paper.}

\end{figure}

\begin{figure}[h!]
\begin{center}
\includegraphics[width=0.75\textwidth]{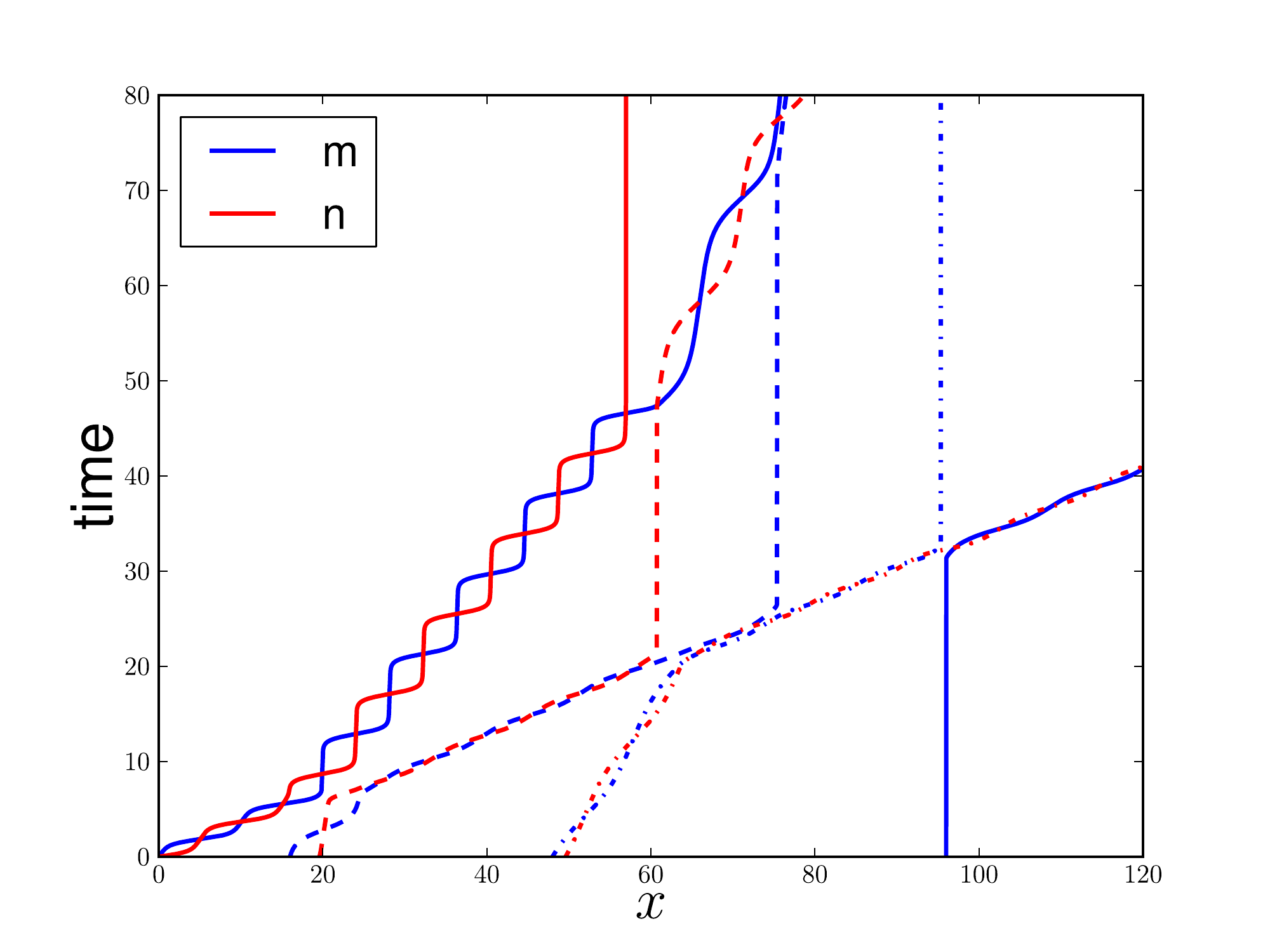}

\caption{\label{fig:mn_compacton_poly_collisions} {\rem{\color{red}\bf Find Waldo on the London Underground.}} Spacetime paths for the problem of sequential pairwise overtaking collisions are plotted for three initially rightward-moving parabolic compacton couples, kernel $K_2$, arranged from left to right in order of decreasing propagation speed and one stationary parabolic compacton monopole. Initial conditions are given in form (species, weight, position) by $(m,8,0)$, $(n,0.8,0)$, $(m,1.6,8)$, $(n,1.6,9.8)$, $(m,0.8,24)$, $(n,0.8,24.8)$ and $(m,1.6,48)$. During the overtaking collisions, the left-right positional order is preserved separately for each species.\\
The first four collisions shown can be classified into three classes:
\begin{enumerate}[(i)]
\item In the first collision, near $t=8$, $x=10$ the two colliding couples remain intact, but suffer sudden changes in their direction of propagation and in the periods of the cycles in both couples. 
\item The second collision at $t=22$, $x=30$, involves the destruction of the faster (dashed) couple, resulting in two unbound monopoles, one of each species, the second of which (dashed blue) is ejected only after propagating for a few cycles as a compound state involving two blues ($m$) and one red ($n$). The collisions at later times all involve exchanges of partners.
\item The penultimate collision again involves formation of a compound state of two blues and one red. The compound state lasts for only two cycles in this case before the exchange is complete and the leftmost one (solid blue) is ejected. 
\item The last collision involves the stationary monopole that was ejected in an earlier exchange collision. It exchanges again and propagates away in a couple. Thus, the initially rightmost positive compacton couples propagate away rightward.
\end{enumerate}
Animations that show the time evolution of this image are available as supplementary material with the online copy of this paper.}

\end{center}
\end{figure}

\paragraph{A new phenomenon for compacton collisions.} The collision problem for parabolic compaction couples with the $K_{2}$ profile does show an additional new property of the compacton systems. 
Namely, pair interactions may force the oscillations past the half-width of the compacton's
support, and thereby separate a waltzing compacton couple into two solitary compactons.
The role of this phenomenon in multi-compacton interactions is explored in order of increasing complexity in Figs. (\ref{fig:mn_compacton_collisions}), (\ref{fig:mn_compacton_poly_collisions}) and (\ref{tube-map-collisions}).

%
%

%

\section{Crossflow equations with more than two species}

As a final example of the application of of this framework, we consider 
equations for multiple (i.e. more than two) species that cannot sustain
 motion without being in each other's presence. For a vector of velocity vector fields, $\bm{u}=\left(u^{(1)},u^{(2)},\ldots,u^{(n)}\right)$, define a Lagrangian
\begin{equation*}
\label{eqn-many-spec-lag}
l\left(\bm{u}\right) = \frac{1}{2}
 (\bm{u},\left(\bm{A}\right)\mathcal{G}\bm{u}),
\end{equation*}
with $\bm{A}$ a real symmetric matrix of determinant $\pm 1$, and $\mathcal{G}$ an arbitrary differential operator, applied elementwise. The resulting Euler-Poincaré equations are 
\begin{equation}
\label{eqn-vec-lag-dt}
\left(\frac{\partial}{\partial t}+\mathcal{L}_{u^{(i)}}\right)\frac{\delta l}{\delta u^{(i)}}=0,
\end{equation}
for a vector of momenta,
$\bm{m}:=\frac{\delta l}{\delta\bm{u}}=\bm{A}\mathcal{G}\bm{u}.$
The inversion of this Legendre transformation is 
\begin{equation}
\label{eqn-vec-Legendre}
\bm{u}=\bm{A}^{-1}\left(K*\bm{m}\right)=K*\bm{A}^{-1}\bm{m},
\end{equation}
where the convolution is also to be applied elementwise. The induced Hamiltonian for this system is thus
\[H=\frac{1}{2}\left(\bm{m},K*\bm{A}^{-1}\bm{m}\right),\]
with Hamiltonian motion equations
\begin{equation}
\label{eqn-vec-ham-dt}
\partial_{t}\bm{m} + 
 \left( \partial_{x}\bm{m} + \bm{m}\partial_{x} \right)
 \frac{\delta H}{\delta\bm{m}} = 0.
\end{equation}

An absence of self-interaction terms in the Lagrangian representation (\ref{eqn-vec-lag-dt}) requires that the matrix $\bm{A}$ have all entries on its leading diagonal vanish. This is the necessary and sufficient sufficient to ensure that no non-linear terms in a single species appear in the governing PDEs for the velocities, meaning initial conditions with non-vanishing velocity of a single species are steady.

 Meanwhile, an absence of self interaction terms in the Hamiltonian representation (\ref{eqn-vec-ham-dt}) requires similarly that the inverse matrix $\bm{A}^{-1}$ have all the leading diagonal vanish, so that the the evolution of each element of $\bm{m}$ is instantaneously independent of its own value.

For $n=2$ the only matrix satisfying our conditions on symmetry, determinant and leading diagonal is 
\[\bm{A}=\left(\begin{array}{cc}
0 & 1\\
1 & 0\end{array}\right),\]
which is unitary and generates generalizations of the CCCH. Thus, for this case of $n=2$ the CCCH are the canonical equations without self-interaction terms. For $n>2$ there may exist systems which are non-self interacting in only one of the two representations. 

As a concrete example, consider the case for $n=3$ and Euler-Poincar\'e Lagrangian 
\[l_{\mathrm{EP}}\left(u^{(1)},u^{(2)},u^{(3)}\right) = 
\frac{1}{4}\int\sum_{j\neq k=1}^{3}u^{(j)}u^{(k)} 
    + u_{x}^{(j)}u_{x}^{(k)}dx,\]
i.e. the coupling matrix for the Euler-Poincar\'e case
\begin{equation}
[\bm{A}^{\mathrm{EP}}]_{ij}=\frac{1}{2}\left(1-\delta_{ij}\right)
\label{EPCMatrix}
\end{equation}
has only  cross-interaction terms.
The choice of coupling matrix $[\bm{A}^{\mathrm{EP}}]$ gives momenta
\begin{equation}
\label{eqn-lag-mom-vel-rel}
m^{(1)} := \frac{\delta l_{\mathrm{EP}}}{\delta u^{(1)}} = 
 \left(1-\frac{\partial^{2}}{\partial x^{2}}\right) 
              \left[\frac{u^{(2)}+u^{(3)}}{2}\right]
\end{equation}
and cyclic permutations of 1,2,3. The velocities then relate to the momenta by
\begin{equation}
u^{(1)}=K*\left[m^{(2)}+m^{(3)}-m^{(1)}\right],
\label{EP-veloc}
\end{equation}
and cyclic permutations of the labels 1,2,3. The EP multi-species cross-coupled system has Hamiltonian
\begin{equation}
H_{\mathrm{EP}}=\frac{1}{2}\int\sum_{j\neq k=1}^{3}m^{(j)}K*m^{(k)}-\sum_{j=1}^{3}m^{(j)}K*m^{(j)}dx,
\label{EP-Ham}
\end{equation}
in which the last term involves self-interaction.

Another, complementary, cross-coupled system exists, in which the \emph{Hamiltonian} representation has no self-interaction terms. This complementary cross-coupled system is a Lie-Poisson Hamiltonian system generated by
\begin{equation}
H_{\mathrm{LP}}=\frac{1}{2}\int\sum_{j\neq k=1}^{3}m^{(j)}K*m^{(k)}dx
\label{LP-Ham}
\end{equation}
with velocities given by, cf. equation (\ref{EP-veloc}),
\begin{equation}
u^{(1)}=K*\left[\frac{m^{(2)}+m^{(3)}}{2}\right]
\label{LP-veloc}
\end{equation}
and cyclic permutations of the labels 1,2,3. 
This formula implies the following Lagrangian for the LP multi-species cross-coupled system,
\[l_{\mathrm{LP}}\left(u^{(1)},u^{(2)},u^{(3)}\right)=\frac{1}{4}\int\sum_{j\neq k=1}^{3}\left[u^{(j)}u^{(k)}+u_{x}^{(j)}u_{x}^{(k)}\right]-\sum_{j=1}^{3}\left[\left(u^{(j)}\right)^{2}+\left(u_{x}^{(j)}\right)^{2}\right]dx\]
and thus a different coupling matrix from (\ref{EPCMatrix}), namely,
\begin{equation}
[\bm{A}^{\mathrm{LP}}]_{ij}=\left[\bm{A}_{\mathrm{EP}}^{-1}\right]_{ij}
=(1-2\delta_{ij})
\label{LPCMatrix}
\end{equation}

\begin{figure}[htp]
\begin{center}

(a)\includegraphics[width=0.45\textwidth]{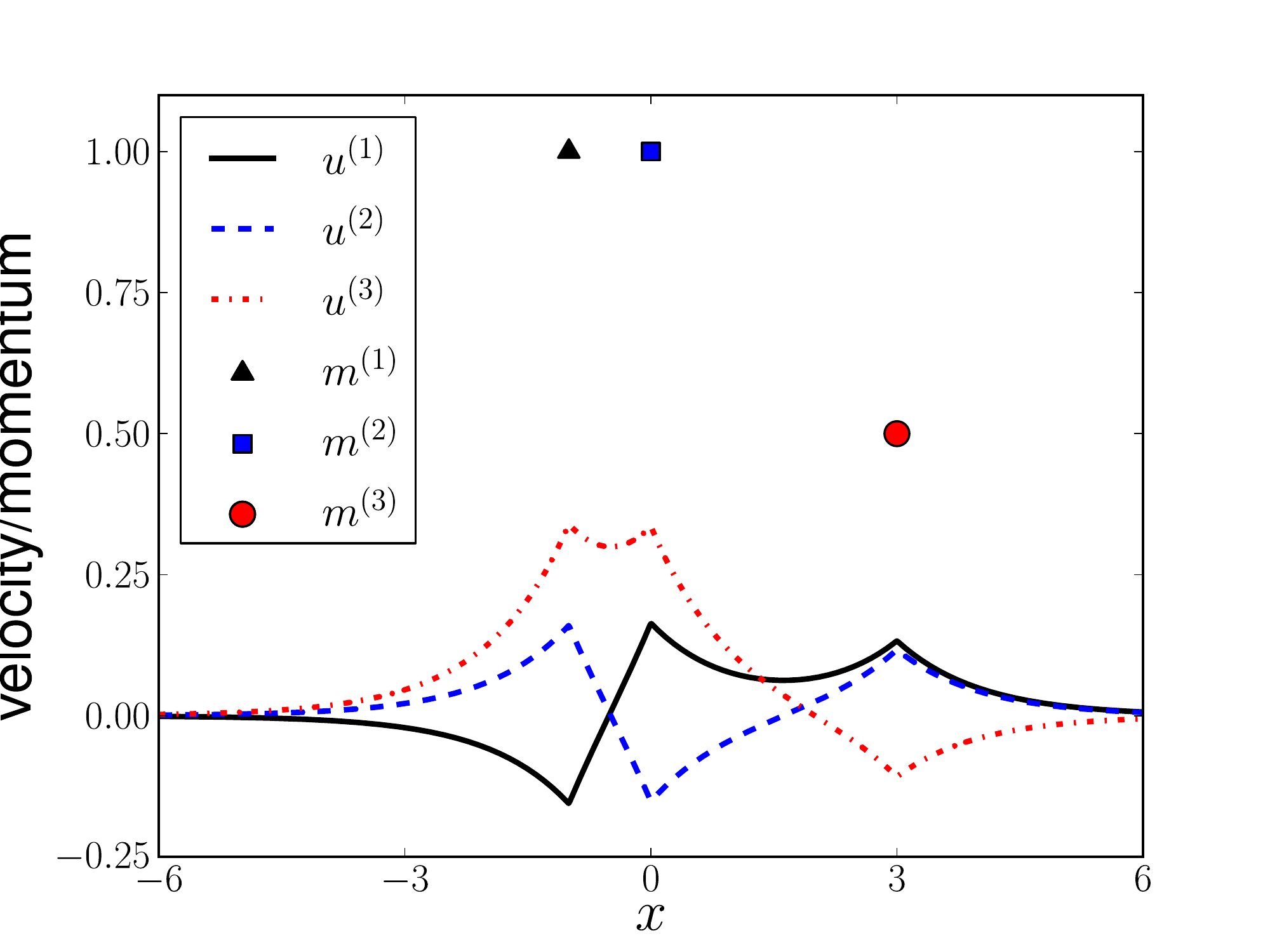}
(b)\includegraphics[width=0.45\textwidth]{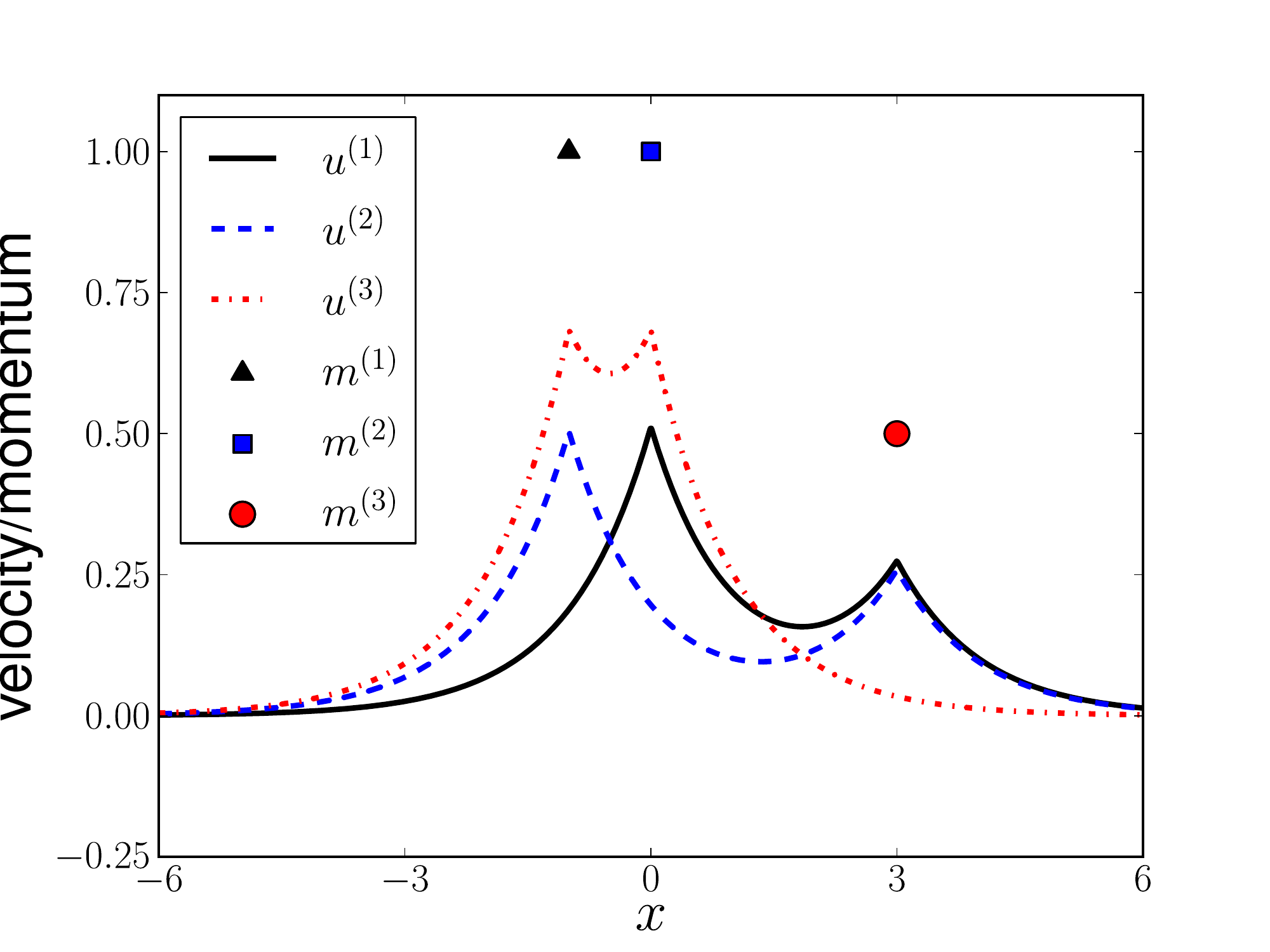}

\caption{\label{fig-vec-ICs} 
Initial conditions for three species cross-coupled equations in the Lagrangian (Panel a) and Hamiltonian representations (Panel b). In each case the locations and weights of the momentum particles, $m^{(i)}$, are identical. However, the resultant velocity fields, $u^{(i)}$, which carry each of the particles differ markedly. Comparision of the panels shows the impact of the strong negative self-interaction term in the viscinity of each particle in the Lagrangian representation.}

\end{center}
\end{figure}

\begin{figure}[htp]
\begin{center}

(a)\includegraphics[width=0.45\textwidth]{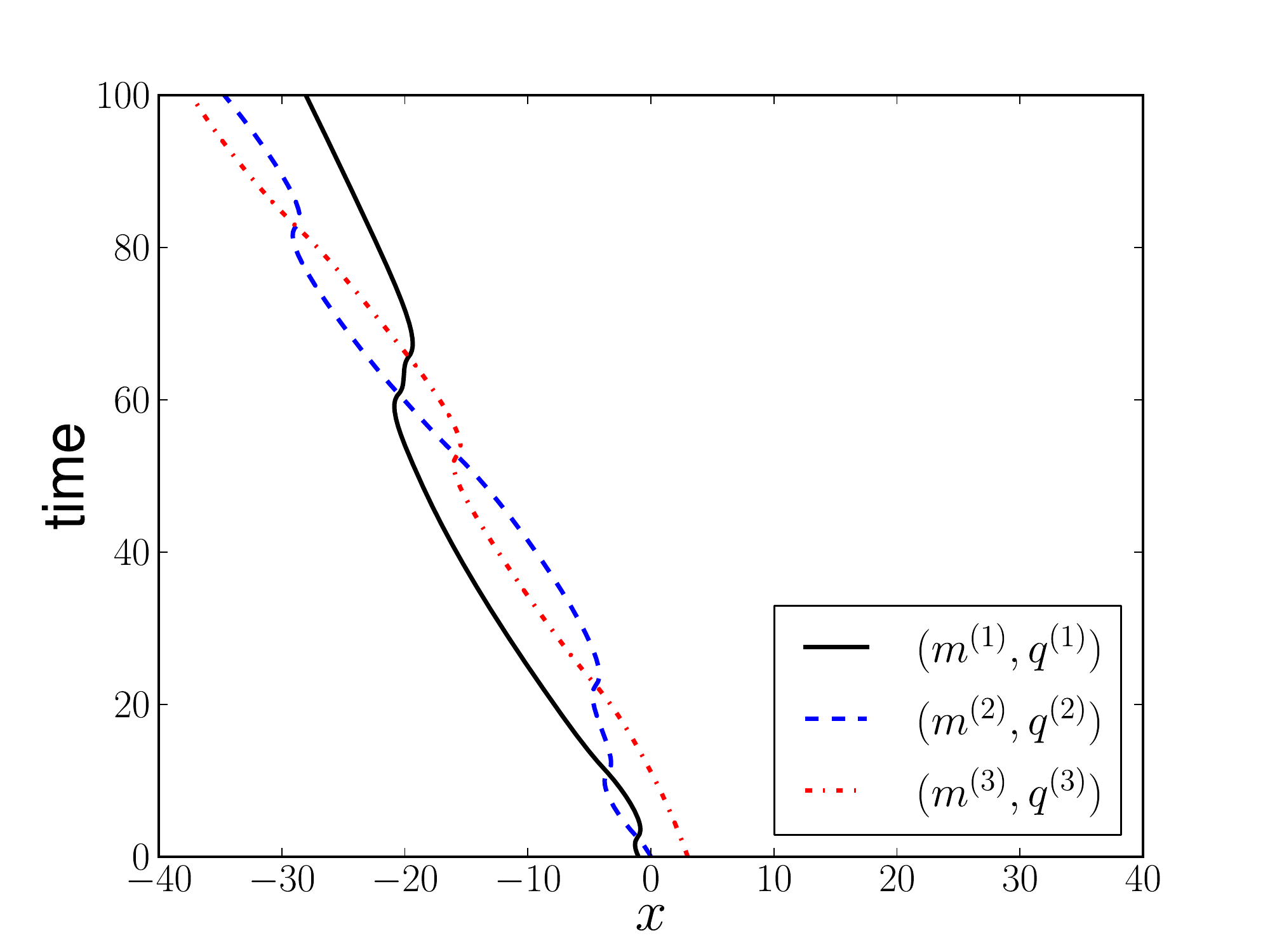}
(b)\includegraphics[width=0.45\textwidth]{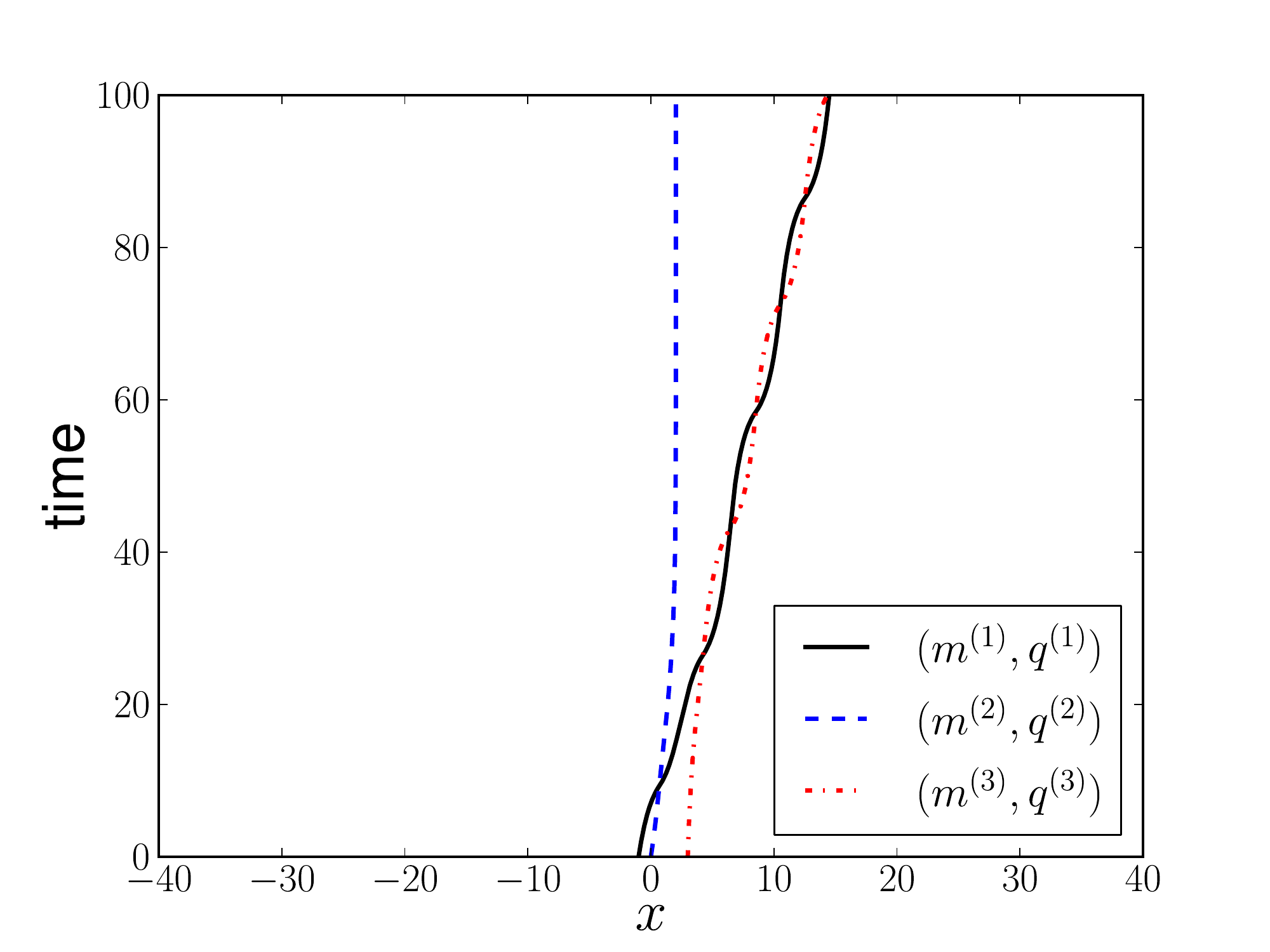}

\caption{\label{fig-vec-spacetime-plots}
Spacetime plots for three species cross-coupled equations in the Lagrangian (Panel a) and Hamiltonian representations (Panel b) with the initial conditions shown in Figure \ref{fig-vec-ICs}.\\
 Panel (a) shows the great significance of the negative momentum self-interaction term which appears in equation (\ref{eqn-lag-mom-vel-rel}). All particles experience a leftward drift, proportional to their own magnitude. Moreover the fully coupled nature of the momentum equations allows.\\
 The Hamiltonian representation in Panel (b) behaves much more in the manner of the CCCH solutions. In fact for these initial conditions the solution is very similar to one in which the $m^{(2)}$ and $m^{(3)}$ particles are of the same species. 
}

\end{center}
\end{figure}

Figures \ref{fig-vec-ICs}--\ref{fig-vec-spacetime-plots} show the results of integrating the singular solutions of these two different equation sets for the same initial conditions (in momentum space). The two equation sets have markedly different behaviour because of their differences in interactions. In particular, the self-interaction term in the EP Hamiltonian (\ref{EP-Ham}) induces a \emph{leftward} drift for positively weighted particles and thereby allows for complexly coupled three-way interactions. In contrast, the three-species problem for the LP cross-coupled equations has fewer differences from CCCH for two species. 

\begin{figure}[h!]
\begin{center}

\includegraphics[width=0.85\textwidth]{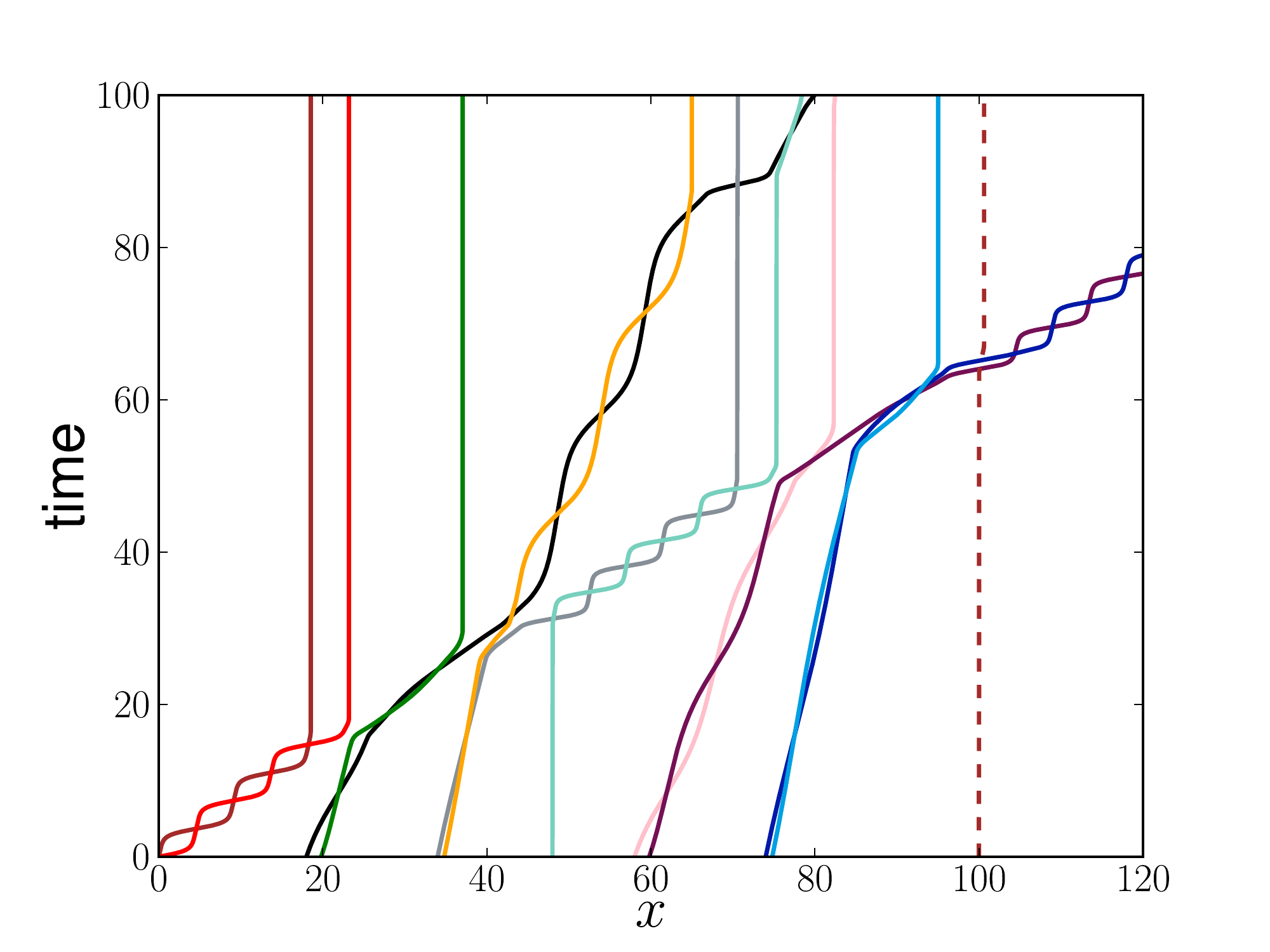}

\caption{\label{tube-map-collisions} 
A space-time plot is shown of the evolution starting from initial conditions consisting of 12 separate species in the LP cross-coupled representation, whose Hamiltonian, as in (\ref{LP-Ham}), has no self-interaction terms. The initial conditions are chosen to form five parabolic compacton couples and two single parabolic compactons. All of the compactons are positive, so all of the couples propagate rightward. The same three basic forms of collision appear as in the previous figures for the CCCH equations however the particles no longer discriminate in their exchange of partners. The final space-time plot is reminiscent of the map of the London Underground!\\
  Animations that show the time evolution of this image are available as supplementary material with the online copy of this paper.}


\end{center}
\end{figure}

As more species are added, the scope for more complex interesting interactions grows. For example, Figure \ref{tube-map-collisions} shows the result of a run with parabolic compactons of \emph{twelve} separate species in the relevant Hamiltonian cross-flow system. This shows both the complex entanglement of the various particles and the trivial classicifaction of many interaction features into those already described in the two species compacton case.

The primary stable features for the multi-species collisions in Figure \ref{tube-map-collisions} are the same as for two species collisions in Figure \ref{fig:mn_compacton_poly_collisions}, in that both include solitary particles and waltzing pairs. However, the transient collision behaviour in the multi-species case can be significantly more complex, possessing, for example, fully entangled three-way collisions, particle tunnelling and ``hop-scotch'', in which a solitary particle is bled of its momentum by a coupled pair, then abandoned.


\section{Conclusions/Discussion}

\label{conclusion-sec}

\paragraph{Explicit outstanding problems.}

Several outstanding problems were identified explicitly in the course
of this work. These include the following tasks.
\begin{enumerate} [(i)] 
\item
Study and classify the full set of interactions of the peakons in (\ref{xflow-peakons}) for the
CCCH system (\ref{cross-flow-eqns}). Determine whether the generalizations
proposed in systems (\ref{cross-flow-eqns}) are integrable, or even whether
the (weak) solutions exist and are unique.
\smallskip

Even if the CCCH equations are not completely integrable, the generalisation they represent is interesting for its singular solutions and geometrical interpretation. In this regard, we may remark about additional symmetry and complete integrability of the corresponding coupled rigid body dynamics. Namely, if the problem specified here for a right-invariant Lagrangian on $\mathfrak{X}(\mathbb{R})\times\mathfrak{X}(\mathbb{R})$ had been expressed instead on $\mathfrak{so}(3)\times\mathfrak{so}(3)$ for a left-invariant Lagrangian, the result would have been interpretable as the dynamics of a cross-coupled system of two rigid bodies. This dynamics would have been expressible in Lie-Poisson form and it would have been completely integrable as a Hamiltonian system, only provided an additional $SO(2)$ symmetry were present, as for the case of axisymmetric coupled rigid bodies. In the present case, there seems to be no extra symmetry that would make the corresponding CCCH Hamiltonian system  (\ref{cross-flow-eqns}) completely integrable.

\item
 Explore the dynamical systems properties of the $(M,N)$-peakon solutions (\ref{xflow-peakons})
of the CCCH equations (\ref{cross-flow-eqns}). For example, determine the results of colliding a compacton couple with a single compacton of opposite sign. \rem{{\color{red}\bf $\Longleftarrow$ Interested in this, James?}}

\item
Characterise the creation of peakons and peakon couples as a process of blow-up in finite time. 
We have sought an analog of the steepening lemma for the CH equation in \cite{CH93} to explain how peakon couples arise from smooth initial conditions for the CCCH equations, but so far this result has eluded us. 

\item
Discuss the solution behaviour of other cross-coupled equations, such as cross-coupled Burgers (CCB), cross-coupled Korteweg de Vries (CCKdV) and cross-coupled $b$-equations. This is a wide open problem. 

\end{enumerate}

\ack

DDH was partially supported by the Royal Society of London, Wolfson
Scheme. RII acknowledges funding from a Marie Curie Intra-European
Fellowship.  JRP was supported by the US ONR. All authors
thank J. Gibbons, C. Tronci, V. Putkaradze for many valuable discussions.
We also thank Y. Brenier for pointing out the similarity of the space-time 
diagram for multiple compacton couple interactions in Figure \ref{fig:mn_compacton_poly_collisions} 
with the map of the London Underground. His comment inspired the creation of 
Figure \ref{tube-map-collisions}.
\\

\end{document}